\RequirePackage{fix-cm}

\documentclass[twocolumn]{svjour3}          

\smartqed 

\usepackage{amsmath,amssymb,amsfonts}
\usepackage[T1]{fontenc}

\usepackage{algorithmic}
\usepackage{graphicx}
\usepackage{textcomp}
\usepackage[dvipsnames]{xcolor}
\def\BibTeX{{\rm B\kern-.05em{\sc i\kern-.025em b}\kern-.08em
    T\kern-.1667em\lower.7ex\hbox{E}\kern-.125emX}}
\usepackage{hyperref}
\usepackage{orcidlink}
\hypersetup{pdfborder={0 0 0}}
\usepackage{listings}
\usepackage{textcomp} 
\usepackage{lstautogobble}
\usepackage[moderate,tracking=normal]{savetrees}
\usepackage[inline]{enumitem}
\setlist[enumerate]{leftmargin=*}

\usepackage[caption=false]{subfig}

\newcommand{\rc}[1]{{\sf {\small #1}}}
\newcommand{\setV}{\mathcal{V}}
\newcommand{\setVars}{\mathbb{V}}
\newcommand{\mkvstate}{\mu}
\newcommand{\pctlOp}[1]{\mathrm{#1}\,}

\usepackage{multirow} %
\usepackage{multicol}
\usepackage{booktabs}

\usepackage{siunitx}
\let\SInum\num 
\let\num\relax

\let\fun\rightarrow
\let\inj\rightarrowtail
\def\surj{\mathrel{\ooalign{$\fun$\hfil\cr$\mkern4mu\fun$}}}
\def\bij{\mathrel{\ooalign{$\inj$\hfil\cr$\mkern5mu\fun$}}}
\def\pfun{\@p\fun}
\def\pinj{\@p\inj}
\def\psurj{\@p\surj}
\def\pbij{\@p\bij}
\def\ffun{\@f\fun}
\def\finj{\@f\inj}

\def\defs{\mathrel{\widehat=}}
\def\union{\mathrel{\cup}}

\definecolor{ClzColor}{cmyk}{0.50,1,0,0} 
\newcommand{\cls}{\color{red}}
\newcommand{\mykeyword}[1]{\emph{#1}}

\usepackage{graphbox} 

\hypersetup{
    colorlinks,
    linkcolor={red!50!black},
    citecolor={blue!50!black},
    urlcolor={blue!80!black}
}

\newcommand{\lstinbnf}[1]{\lstinline[language=AssertBNF,columns=fullflexible,breaklines=true]{#1}}
\newcommand{\lstinrbc}[1]{\lstinline[language=RoboCert,columns=fullflexible,breaklines=true]{#1}}
\newcommand{\lstinxtd}[1]{\lstinline[language=Xtend,columns=fullflexible,breaklines=true]{#1}}
\newcommand{\lstinxtx}[1]{\lstinline[language=Xtext,columns=fullflexible,breaklines=true]{#1}}

\lstdefinelanguage{PRISM}{
  morekeywords={
      dtmc, const, int, float, bool, global, module, init, endmodule, false, double,
      rewards, endrewards, formula
  },
  sensitive=true, 
  morecomment=[l]{//}, 
  morecomment=[is]{/*}{*/}, 
  morestring=[b]" 
} %

\definecolor{eclipseBlue}{RGB}{42,0.0,255}
\definecolor{eclipseGreen}{RGB}{63,127,95}
\definecolor{eclipsePurple}{RGB}{127,0,85}
 
\lstdefinelanguage{RoboCert}{
  morekeywords={
    csp-begin, csp-end, csp, untimed, CHAOS, RUN, Events,
    in, out, 
  },
  alsoletter=-,
  sensitive=true, 
  morecomment=[l]{//}, 
  morecomment=[is]{/*}{*/}, 
  morestring=[b]" 
} %

\lstdefinelanguage{AssertBNF}{
  morekeywords={
    in, out, val, 
    import, package,  
    prob, with, cmdoptions, property,
    const, constant, constants, and, set, to, assigned, value, with, from, set,
    label, formula,
    rewards, endrewards, 
    definitions, defs, pfunction, poperation, return,
    Prob, of, Forall, Exists, min, max, 
    Next, Until, Finally, Globally, Weak, Release,
    Reachable, Cumul, LTL, Total,
    using, sim, pathlen, CI, ACI, APMC, SPRT,   
    alpha,
    if, then, else, end, Not, is, in, true, false,
    modules, pmodules, pmodule, skip, bool, init, to,
    not, iff,
    by, step,
    init, deadlock,
    epsilon, delta, 
  },
  sensitive=true, 
  morecomment=[l]{//}, 
  morecomment=[is]{/*}{*/}, 
  morestring=[b]", 
  morestring=[b]', 
} %

\definecolor{lightgray}{rgb}{.9, .9, .9}
\definecolor{darkgray}{rgb}{.4, .4, .4}
\definecolor{purple}{rgb}{0.65, 0.12, 0.82}

\lstdefinelanguage{Epsilon}{
  morekeywords={
      if, then, else, operation, var, and, or, not, new,
      @lazy, rule, transform, to, self
  },
  sensitive=true, 
  morecomment=[l]{//}, 
  morecomment=[is]{/*}{*/}, 
  morestring=[b]" 
} %

\definecolor{epsilonred}{rgb}{0.6,0,0} 
\definecolor{epsilongreen}{rgb}{0.25,0.5,0.35} 
\definecolor{epsilonpurple}{rgb}{0.5,0,0.35} 
\definecolor{epsilondocblue}{rgb}{0.25,0.35,0.75}

\lstdefinelanguage{Xtend}{%
  morekeywords={%
    cached,%
    case,%
    default,%
    extension,%
    false,%
    import,%
    JAVA,%
    WORKFLOWSLOT,%
    let,%
    new,%
    null,%
    private,%
    create,%
    switch,%
    this,%
    true,%
    reexport,%
    around,%
    if,%
    then,%
    else,%
    def,%
    @Check,%
    val,%
    var,%
    as,%
    instanceof,%
    return,%
    override,
    for,
    IF,
    ENDIF,
  },
  keywordstyle=[2]{\textbf},%
  morecomment=[l]{//},%
  morecomment=[s]{/*}{*/},%
  morestring=[b]",%
  tabsize=4%
}%

\lstdefinelanguage{Xtext}{%
  morekeywords={%
    returns, current%
  },
  keywordstyle=[2]{\textbf},%
  morecomment=[l]{//},%
  morecomment=[s]{/*}{*/},%
  morestring=[b]",%
  tabsize=4%
}%

\begin{document}

\title{RoboCertProb: Property Specification for Probabilistic RoboChart Models}
\titlerunning{RoboCertProb: Property Specification for Probabilistic RoboChart models}

\author{Kangfeng Ye \and Jim Woodcock}
\institute{University of York, York, UK \\ \email{\{kangfeng.ye,jim.woodcock\}@york.ac.uk}}

\authorrunning{Kangfeng Ye \and Jim Woodcock}

\maketitle

\begin{abstract}
  RoboChart is a core notation in the RoboStar framework which brings modern modelling and formal verification technologies into software engineering for robotics. It is a timed and probabilistic domain-specific language for robotics and provides a UML-like architectural and state machine modelling. This work presents RoboCertProb for specifying quantitative properties of probabilistic robotic systems modelled in RoboChart.
  RoboCertProb's semantics is based on PCTL*. To interpret RoboCertProb over RoboChart models, we give a Markov semantics (DTMCs and MDPs) to RoboChart, derived from its existing transformation semantics to the PRISM language. In addition to property specification, RoboCertProb also entitles us to configure loose constants and unspecified functions and operations in RoboChart models. It allows us to set up environmental inputs to verify reactive probabilistic systems not directly supported in probabilistic model checkers like PRISM because they employ a closed-world assumption.
  We implement RoboCertProb in an accompanying tool of RoboChart, RoboTool, for specifying properties and automatically generating PRISM properties from them to formally verify RoboChart models using PRISM. We have used it to analyse the behaviour of software controllers for two real robots: an industrial painting robot and an agricultural robot for treating plants with UV lights.
\end{abstract}

\section{Introduction}

RoboChart~\cite{Miyazawa2019,Ye2022} is a core notation in the RoboStar%
\footnote{%
  \url{robostar.cs.york.ac.uk}.%
} %
framework~\cite{Cavalcanti2021} brings modern modelling and verification technologies into software engineering for robotics to address the following challenges in the current practice of programming robotic applications:
\begin{enumerate*}[label={(\alph*)}]
\item no precise syntax and formal semantics, 
\item informally discussed time and uncertainty requirements, 
\item loosely connected artefacts, 
\item no tool support, and 
\item no assurance.
\end{enumerate*}
In this framework, platform, environment, design, and simulations are models with formal mathematical semantics in a unified semantic framework~\cite{Hoare1998}. Modelling, semantics generation, verification, simulation, and testing are automated and integrated into an Eclipse-based tool, RoboTool.%
\footnote{\url{robostar.cs.york.ac.uk/robotool/}.}

RoboChart is a UML-like architectural and state machine modelling notation featuring discrete time and probabilistic modelling to deal with environmental uncertainty, such as an unknown map. The formal semantics~\cite{Miyazawa2019} for its standard (non-probabilistic) state machines and time features is based on the CSP process algebra~\cite{Hoare1985,Roscoe2011} and the semantics~\cite{Woodcock2019,Ye2022} for its probabilistic feature is based on probabilistic designs~\cite{Ye2021} in Hoare and He's Unifying Theories of Programming (UTP)~\cite{Hoare1998} and the PRISM language~\cite{Kwiatkowska2011}. 

RoboCert~\cite{Ye2022,Windsor2022} is a notation to specify properties for RoboChart models. It allows users to specify qualitative properties of non-probabilistic RoboChart models in sequence diagrams, presented in~\cite{Windsor2022}, and both qualitative and quantitative properties of probabilistic RoboChart models in temporal logics, briefly discussed in~\cite{Ye2022} and presented here for a complete account. In this paper, we refer to the work~\cite{Windsor2022} as RoboCertSeq (to avoid confusion), which is based on UML sequence diagrams and has its semantics in \emph{tock}-CSP~\cite{BaxterT21,Roscoe2011}. Here, we present the probabilistic counterpart of RoboCertSeq, called RoboCertProb.
It is not an extension of RoboCertSeq
and has different semantics. RoboCertProb is based on PCTL*~\cite{Aziz1995,Bianco1995,Baier1998}, a combination of the probabilistic computation tree logic (PCTL)~\cite{Hansson1994} and the linear temporal logic (LTL)~\cite{Pnueli1977}, or seen as the probabilistic counterpart of CTL*~\cite{Emerson1986} combining CTL~\cite{Clarke1982} and LTL.

PCTL* is interpreted over discrete Markov models, such as Discrete-Time Markov Chains (DTMCs)~\cite{Kemeny1976} and Markov Decision Processes (MDPs)~\cite{Howard1971,Puterman1994}. The semantics for RoboChart models given in~\cite{Ye2022} is based on the transformation to models in the PRISM language. Then the PRISM model checker will compile and build the PRISM models into underlying DTMCs or MDPs according to the specified Markov model type in the PRISM models, as stated in the PRISM semantics~\cite{PRISMTeam2008}. That the semantics of RoboChart models is in PRISM, indeed, facilities the development of RoboTool to automatically generate PRISM models using model-based transformation. This, however, makes it impossible to interpret PCTL* over RoboChart models directly because of the lack of the semantics in DTMCs and MDPs. For this reason, we give RoboChart models the semantics in DTMCs and MDPs directly in this paper. We note that this Markov semantics is consistent with the semantics in PRISM because it is derived from the PRISM semantics. The Markov semantics for RoboChart models enables us to specify properties using PCTL*. 

The use of PCTL* directly to specify properties, however, is subject to several problems. 
Firstly, it is against our intention to design domain-specific languages (DSLs) for roboticists. Secondly, it is a challenge to choose the right states or transitions to specify for our properties in the resultant Markov semantics because the component structure of RoboChart models will be significantly changed (flattened) and additional states are introduced in the Markov semantics for this purpose. Thirdly, the expressions in PCTL* to identify states based on their associated atomic propositions (APs) are simple, which makes it inconvenient to quantify multiple states. Instead, we can use predicates to facilitate specifications. For this reason, we need a DSL property specification language with rich expressions, integrated with RoboChart models seamlessly. This motivates the design of RoboCertProb.

In addition to specifying properties, RoboCertProb aims to configure loose constants and define unspecified functions and operations in RoboChart models for verification based on the model instantiations. A translation of RoboChart models to PRISM is presented in~\cite{Ye2022}, and then probabilistic model checking using PRISM is conducted on the generated PRISM models. RoboChart models reactive robotic systems. PRISM, however, employs a closed-world assumption: systems are not subjected to environmental inputs.
RoboCertProb introduces a feature to specify environmental inputs and check outputs from systems in additional PRISM modules in parallel composition with the generated PRISM models to verify reactive systems in RoboChart. Using the reachability checking in PRISM, we can achieve a similar trace refinement checking of the RoboChart models using FDR~\cite{T.GibsonRobinson2014} to verify safety properties.


We favour a controlled natural language (CNL) syntax for RoboCertProb. Its syntax is very rich and flexible.
We implemented RoboCertProb in RoboTool to support modelling, validation of well-formedness conditions, and code generation of properties written in RoboCertProb to PRISM properties.

Our novel contributions are as follows: 
\begin{enumerate*}[label={(\alph*)}]
\item the Markov semantics for RoboChart models in both DTMCs and MDPs, 
\item a PCTL*-based property specification RoboCertProb (a CNL) for RoboChart, which allows model instantiations, model references, environment modelling, and quantitative measurement;
\item the implementation of RoboCertProb as a plug-in of RoboTool; and
\item the use of RoboCertProb to verify RoboChart models for an industrial painting robot and an agricultural UV-light treatment robot.
\end{enumerate*}
Though this work is specific to RoboChart, we believe the methodology and techniques can be generalised mainly to other domain-specific languages to address the challenges in adapting formal specification and verification for wider users (like roboticists) with little knowledge of formal methods.

The remainder of this paper is organised as follows. We introduce RoboChart by modelling the software controller for a bounded random walk in Sect.~\ref{sec:robochart}. This is the running example in this paper to illustrate our approach. In Sect.~\ref{sec:markov_semantics}, we then present our Markov semantics for RoboChart, the semantics of PCTL*, and our motivations for the design of RoboCertProb, with the illustrations for the running example. Section~\ref{sec:ppl_syntax} details RoboCertProb from its syntax, well-formedness conditions, design decisions, and illustrations with examples. We then describe the detailed implementation of RoboCertProb in RoboTool to support modelling, validation, and automatic generation in Sect.~\ref{sec:tool} and discuss the verification of three robotic examples using RoboCertProb in Sect.~\ref{sec:examples}. Finally, we review related work in Sect.~\ref{sec:related} and conclude in Sect.~\ref{sec:conclusion}.

\section{RoboChart}
\label{sec:robochart}

RoboChart has a component model with notions of controller and module, and state machines to foster reuse.  RoboChart's metamodel determines the structure of RoboChart models and specifies the constructs' types and relations. We refer to the RoboChart reference manual~\cite{A.Miyazawa2020} for a complete account of RoboChart metamodel.

We describe the facilities of RoboChart for modelling probabilistic controllers of robots using an example of a simple bounded random walk (SRW). This robot walks along a one-dimensional line and starts from its origin at position 0. This robot also has a randomisation device, such as a coin flip. Before every movement, it flips a coin. If the outcome is ``heads'', the robot moves one step to its right, increasing its position by 1.  If the outcome is tails, the robot moves one step to its left, decreasing its position by 1. We note that the space is bounded by the maximum distance \rc{MaxDist} the robot can move away from the origin, and the number of steps the robot can move is also bounded by the maximum allowed steps \rc{MaxSteps}.

\begin{figure*}[tb]
  \centering
  \includegraphics[scale=0.60]{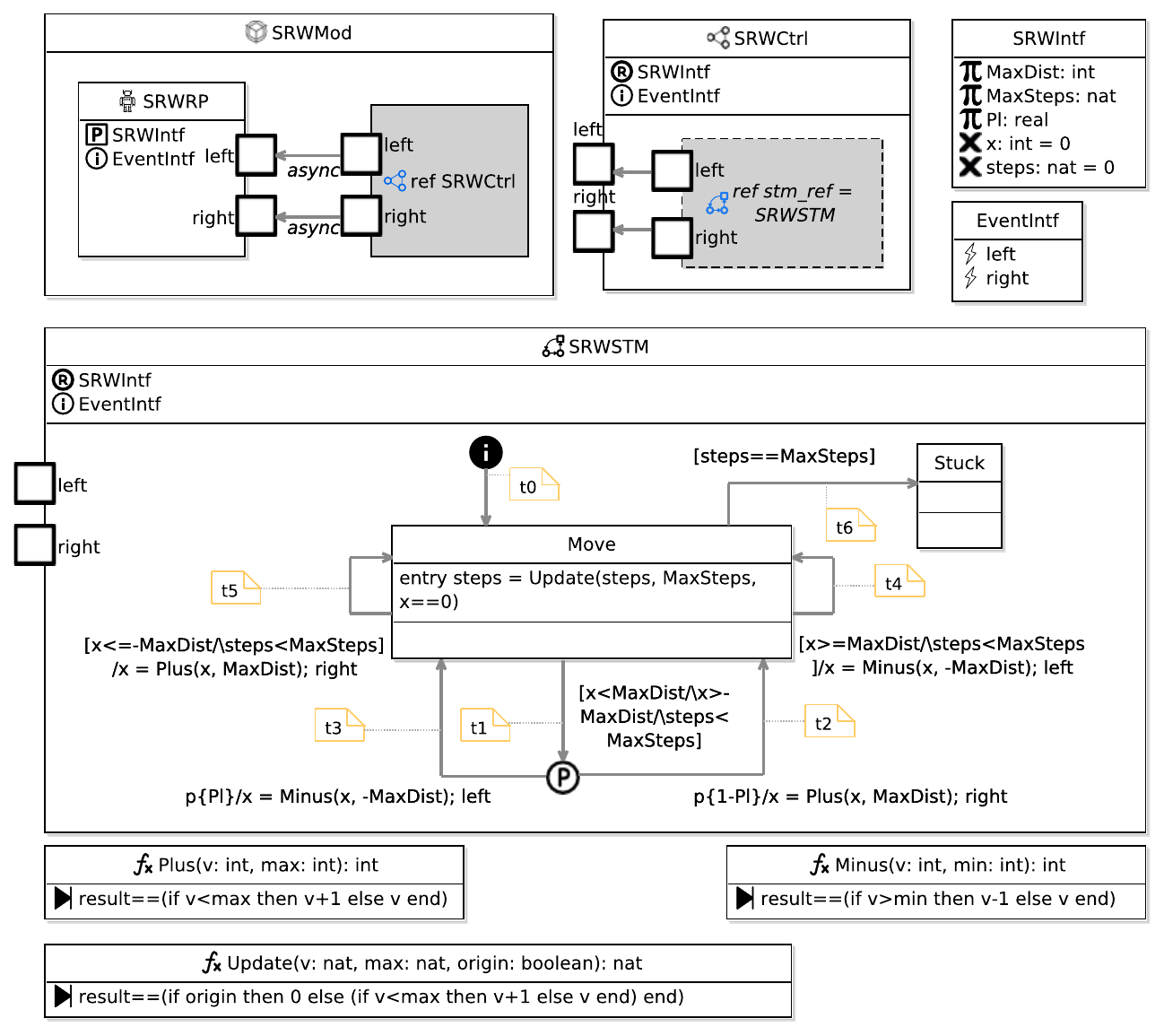}
  \caption{The RoboChart model for the simple random walk: data interface~\rc{SRWIntf}, event interface~\rc{EventIntf}, module~\rc{SRWMod}, robotic platform~\rc{SRWRP}, controller~\rc{SRWCtrl}, state machine~\rc{SRWSTM}, and three functions~\rc{Plus}, \rc{Minus}, and \rc{Update}.}
  \vspace*{-1em}
  \label{fig:srw_model}
\end{figure*}
The RoboChart model of the controller of this robot is presented in Fig.~\ref{fig:srw_model}.  The model is a module \rc{SRWMod} containing a robotic platform \rc{SRWRP} and a controller \rc{SRWCtrl} (through a controller reference). The controller includes a state machine \rc{SRWSTM} (though a state machine reference).


Physical robots are abstracted into robotic platforms through variables, events, and operations. The platform \rc{SRWRP} provides (\includegraphics[align=c,height=8pt]{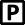}) various shared variables and constants through an interface \rc{SRWIntf}, which is required (\includegraphics[align=c,height=8pt]{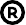}) by the controller and the state machine. There are three constant variables: \rc{MaxDist} of type \rc{int} (integer), \rc{MaxSteps} of type \rc{nat} (natural numbers), and \rc{Pl} of type \rc{real} (real numbers) denoting the probability of the coin flip being heads (and so move right); and two variables: \rc{x} of type \rc{int} for the current position of the robot, and \rc{steps} of type \rc{nat} to record how many steps the robot has moved. Both variables are initialised to 0.

In addition to the shared variables, the platform \rc{SRWRP}, the controller \rc{SRWCtrl}, and the machine \rc{SRWSTM} communicate using directional connections on two events (\includegraphics[align=c,height=8pt]{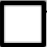}) \rc{left} and \rc{right} (which are defined through an interface \rc{EventIntf}). The controller relays the events from the machine to the platform.

The behaviour of this model is captured in the state machine \rc{SRWSTM}, which contains four nodes: one initial junction (\includegraphics[align=c,height=8pt]{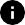}), a probabilistic junction (\includegraphics[align=c,height=8pt]{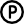}), and two states \rc{Move} and \rc{Stuck}. The \rc{Move} has an entry action, \rc{steps=Update(steps,MaxSteps,x==0)}, to update \rc{steps} using a function \rc{Update} specified by a postcondition (\includegraphics[align=c,height=8pt]{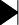}). This action resets \rc{steps} to 0 if the robot is back to the origin, increases \rc{steps} by 1 if the maximum steps \rc{MaxSteps} has not reached, and does not change \rc{steps} otherwise. 
RoboChart supports several action constructs: \rc{skip}, an action that terminates immediately, assignment (=), sequential composition (;), conditional (\rc{if}), and communication including input events (of the form e?x where e is an event and x is a variable that records the input value), output events (of the form e!v where v is an expression whose value is output), or synchronisations (of the form e or e.v). The entry action above is an assignment. Other actions in the model such as \rc{x=Plus(x,MaxDist);right}, use assignment, sequential composition, and synchronisation constructs. The constructs \rc{skip}, assignment, and communication are \mykeyword{atomic} (that is, as long as they starts, they terminates immediately without interruption). Sequential composition is non-atomic. The composition of atomic actions by conditional \rc{if} is atomic, and otherwise, it is non-atomic. 

The nodes are connected by seven transitions annotated as \rc{t0} to \rc{t6}.  Transitions have a label with the following optional features: a \emph{trigger} event, a \emph{guard} specifying the conditions that need to hold for the transition to be enabled, a \emph{probability} value defining the probability of occurrence of this transition, and an \emph{action} that is executed if the transition is taken.  The machine \rc{SRWSTM} starts with its initial junction and enters \rc{Move} after its default transition \rc{t0} is taken. Depending on the value of \rc{x} and \rc{steps}, one of the transitions \rc{t1}, \rc{t4}, \rc{t5}, and \rc{t6} is enabled and can be taken. If \rc{steps} is equal to \rc{MaxSteps}, only \rc{t6} is enabled. If it is taken, the machine leaves \rc{Move} and enters \rc{Stuck}. Since \rc{Stuck} has no outgoing transitions, the machine will stay in this state permanently.  If \rc{steps} is less than \rc{MaxSteps}, and the robot has not reached the maximum distance from the origin, that is, \rc{x$<$MaxDist} and \rc{x$>$-MaxDist}, \rc{t1} is enabled and changes the machine from \rc{Move} to the probabilistic junction where a choice is made probabilistically based on the value of \rc{Pl}: \rc{t3} to move to its \rc{left} with probability \rc{Pl} and \rc{t2} to its \rc{right} with probability \rc{1-Pl}. Accordingly, \rc{t3} and \rc{t2} will change the position \rc{x} by the functions \rc{Plus} and \rc{Minus} defined for the bounded increase and decrease of $x$ according to \rc{MaxDist} and \rc{-MaxDist}. If the robot has reached the maximum distance on the right, that is, \rc{x $\geq$ MaxDist}, the self-transition \rc{t4} is enabled to move to its \rc{left} and decrease $x$ by 1 using \rc{Minus}. Similarly, if the robot has reached the maximum distance on the left, that is, \rc{x $\leq$ -MaxDist}, \rc{t5} is enabled to move to its \rc{right} and increase $x$ by 1 using \rc{Plus}.

\section{Markov semantics for RoboChart and PCTL* for specification}
\label{sec:markov_semantics}
\subsection{Markov models: DTMCs and MDPs}
\label{sec:markov_definitions}
We consider two Markov models for RoboChart models: Discrete-time Markov Chains (DTMCs)~\cite{Kemeny1976} and Markov Decision Processes (MDPs)~\cite{Howard1971,Puterman1994}. 

%

A DTMC is a sequence of discrete random variables $X_1, X_2, X_3, \ldots$ with the \emph{Markov property} (that is, the probability of the future system states $s_{n+1}$ depends only on its current state $s_n$, and is independent of all its past states). The possible values of the $X_i$s form a countable set $S$ called its state space. The formal DTMC definition is given below.
\begin{definition}[Discrete-Time Markov chains]
  \label{def:dtmc}
  A DTMC is a tuple $\left(S, s_{init}, P, L\right)$ where
  \begin{itemize}
  \item $S$ is a non-empty and countable set of states;
  \item $s_{init} \in S$ is an initial state; 
  \item $P: S \times S \fun [0,1]$ is a transition probability matrix such that $\sum\limits_{s' \in S}^{} P(s, s') = 1$ for all $s \in S$; 
  \item $L: S \fun 2 ^ {AP}$ is a labelling function where $AP$ denotes a set of atomic propositions.
  \end{itemize}   
\end{definition}

States and transitions of a DTMC can be associated with rewards or cost.
\begin{definition}[Cost and rewards]
    A state reward function in DTMCs is given by $R_s: S \fun \mathcal{R}_{\geq 0}$,  associating a state with a non-negative real number reward. Similarly, a transition reward function $R_t: S \times S \fun \mathcal{R}_{\geq 0}$ associates a transition with a reward. $R_t$ is also called a transition reward matrix.
\end{definition}

MDPs are an extension of DTMCs to allow nondeterministic choice. A formal definition of a MDP is given as follows.
\begin{definition}[Markov decision processes]
  \label{def:mdp}
  A MDP is a tuple $\left(S, s_{init}, Act, Steps, L\right)$ where
  \begin{itemize}
  \item $S$ is a non-empty and countable set of states;
  \item $s_{init} \in S$ is an initial states; 
  \item $Act$ is a set of actions;
  \item $Steps: S \times Act \times S \fun [0,1]$ is a transition probability function such that $\sum\limits_{s' \in S}^{} Steps(s, \alpha, s') \in \{0, 1\}$ for all $s \in S$ and $\alpha \in Act$.
  \item $L: S \fun 2 ^ {AP}$ is a labelling function.
  \end{itemize}   
\end{definition}
Unlike DTMCs where $P(s)$ for each state $s$ is a distribution (the probabilities for all its target states sum to 1), $Steps(s)$ in an MDP is a set of distributions which is indexed by an action $\alpha$. We note that every transition in a DTMC or MDP takes one unit of (discrete) time.

\subsection{Markov semantics for RoboChart}
\label{sec:markov:robochart}
In \cite{Ye2022}, we give RoboChart models the Markov semantics (DTMCs and MDPs) in PRISM directly. A \mykeyword{state} in Markov models is a valuation of all variables in the corresponding PRISM model. A \mykeyword{partial state} is a valuation of partial variables. Here we give RoboChart the Markov semantics in the definitions of DTMCs and MDPs as shown in Definitions~\ref{def:dtmc} and~\ref{def:mdp} to facilitate the definition of semantics for our property language RoboCertProb.

First, we define a corresponding state $\mkvstate$ in Markov models, called a \emph{Markov state} shown in Fig.~\ref{fig:semantics:markov_state}, to a RoboChart model. 
\begin{figure*}
    \centering
\begin{align*}
    \mkvstate = & \left(\setVars, C_1.\setV, \cdots, C_n.\setV\right) \tag*{(1)} \label{eqn:1}\\
    = & \left(\setVars, \overbrace{\left(\setVars_{c_1}, M_1.\setV, \cdots, M_m.\setV\right)}^{C_1.\setV}, \cdots, \overbrace{\left(\setVars_{c_n}, M_1.\setV, \cdots, M_l.\setV\right)}^{C_n.\setV} \right) \tag*{(2)} \label{eqn:2}\\
    = & \left(\setVars, \overbrace{\left(\setVars_{c_1}, \underbrace{\left(\setVars_{m_1}, lk, CS_1.\setV,\cdots,CS_p.\setV\right)}_{M_1.\setV}, \cdots, M_m.\setV\right)}^{C_1.\setV}, \cdots, \overbrace{\left(\setVars_{c_n}, M_1.\setV, \cdots, M_l.\setV\right)}^{C_n.\setV} \right)\tag*{(3)} \label{eqn:3} \\
   = & \left(\setVars, \overbrace{\left(\setVars_{c_1}, \overbrace{\left(\setVars_{m_1}, lk, \underbrace{\left(pc, exit\right)}_{CS_1.\setV},\cdots,CS_p.\setV\right)}^{M_1.\setV}, \cdots, M_m.\setV\right)}^{C_1.\setV}, \cdots, \overbrace{\left(\setVars_{c_n}, M_1.\setV, \cdots, M_l.\setV\right)}^{C_n.\setV} \right) \tag*{(4)} \label{eqn:4} 
\end{align*}
    \caption{Definition of a Markov state.}
    \label{fig:semantics:markov_state}
\end{figure*}

A state $\mkvstate$ is a tuple \ref{eqn:1}, composed of the set $\setVars$ of shared variables provided by the robotic platform, and partial states $C_i.\setV$ for each controller $C_i$. A $C_i.\setV$ further contains the set $\setVars_{c_i}$ of the controller defined variables, and partial states $M_i.\setV$ for each state machine $M_i$, as shown in~\ref{eqn:2}. Similarly, a $M_i.\setV$ contains the set $\setVars_{m_i}$ of the machine defined variables, a $lk$ lock variable, and partial states $CS_i.\setV$ for each composite state $CS_i$, as shown in~\ref{eqn:3}. A $CS_i.\setV$ contains a $pc$ program counter variable and an $exit$ variable to deal with the exit from a RoboChart state, as shown in~\ref{eqn:4}. We refer to~\cite{Ye2022} for more details about the definition and usage of $lk$, $pc$, and $exit$ variables.

The Markov states of the RoboChart model in Fig.~\ref{fig:srw_model} are the valuations of $\left(x, steps, \left(\left(lk, \left(pc, exit\right)\right)\right)\right)$, short as $\left(x, steps, lk, pc, exit\right)$ after the removal of several parenthesis for this simple model. We can also remove $exit$ because it is introduced for composite states and this simple model only contains simple states. One of such states, for example, would be $\left(0, 0, 0, i0 \right)$, denoting its initial state $\mkvstate_{init}$ where $x=0$, $steps=0$, $lk=0$ (that is, the lock is free and no transition is taken), and $pc=i0$ (that is, the state machine is at its initial junction \rc{i0}).

Each state in Markov models is labelled with a set of APs. We define a \emph{labelling function} $L$ for each state as a set of valuations, one valuation for each component in the state. $L$ is defined based on the labelling functions $L_{cs}$ for a composite state, $L_m$ for a state machine, and $L_C$ for a controller.
\begin{align*}
    & L_{cs}\left(CS.\setV\right) = L_{cs}\left(\left(pc=c_s,exit=c_e\right)\right) \defs \left\{pc=c_s,\ exit=c_e\right\}\\ 
    & L_m\left(M.\setV\right) = L_m\left(\left(\setVars_{m}=c_v, lk=c_l, {CS_1.\setV},\cdots,CS_p.\setV\right)\right) \defs \\
    & \quad \left\{v:M.\setVars_{m} \bullet v=c_v(v) \right\} \union \left\{lk=c_l\right\} \union \bigcup \left\{cs:\mathbb{CS}.\setV \bullet L_{cs}(cs)\right\} \\ 
    & L_C\left(C.\setV\right) = L_C\left(\left(\setVars_{c}=c_v, {M_1.\setV},\cdots,M_m.\setV\right)\right) \defs \\
    & \quad \left\{v:C.\setVars_{m} \bullet v=c_v(v) \right\} \union \bigcup \left\{m:\mathbb{M}.\setV \bullet L_{m}(m)\right\} \\ 
    & L(\mkvstate) \defs \left\{v:\setVars \bullet v=c_v(v) \right\} \union \bigcup\left\{c:\mathbb{C} \bullet L_C(c)\right\}
\end{align*}
The partial state $CS.\setV$ for each composite state $CS$ is labelled (by $L_{CS}$) with two APs: a valuation $pc=c_s$ of $pc$ as a value $c_s$ and a valuation $exit=c_e$ of $exit$ as a value $c_e$. We note that a composite state $CS$ can further contain other composite substates and so APs for these substates too. We omit these details here for simplicity.

The partial state $M.\setV$ for a state machine $M$ is labelled (by $L_M$) with three sets of APs. The first set corresponds to the valuations for all state machine defined variables $M.\setVars_m$, in which each variable $v$ takes a value $c_v(v)$ where $c_v$ is a function mapping variables to expressions. The second set contains only one element, the valuation of variables $lk$ as $c_l$. And the third set is a generalised union ($\bigcup$) of the labelling APs for all composite states $\mathbb{CS}$ by $L_{cs}$. 

The labelling function $L_C$ for a partial state $C.\setV$ of a controller $C$, is similar to $L_m$ except that there is no set related to $lk$ and $C$ contains machines $\mathbb{M}$, whose labelling function is $L_m$.

Finally, the labelling $L(\mkvstate)$ of a Markov state $\mkvstate$ is the union of the labelling APs for the valuations of share variables $\setVars$ and APs for all controllers $\mathbb{C}$. For example, $L(\mkvstate_{init})$ for the random walk example is \[\left\{(x=0), (steps=0), (lk=0), (pc=i0)\right\}\] which contains four elements (of which each element is a valuation of one variable in its state).

After the introduction of Markov states and the labelling function for RoboChart models, we define \emph{Markov transitions} as follows. 
First, we introduce a notation below to denote transitions in RoboChart. 
\begin{align*}
    t = e[g]\langle a_x\rangle \langle \left\{p_1/a_{t_1} \to \lfloor a_{e_1}\rfloor s_1\right\} + \dots + \left\{p_n/a_{t_n} \to \lfloor a_{e_n} \rfloor s_n\right\} \rangle
\end{align*}
This denotes a transition ${t}$ from source state ${s}$, with trigger ${e}$ and guard condition ${g}$, to $n$ target states ($s_1,\dots,s_n$) based on their corresponding probabilities ($p_1,\dots,p_n$). Additionally, the exit action of ${s}$ is ${a_x}$, the transition action to each target state $s_i$ is $a_{t_i}$, and the entry action to $s_i$ is $a_{e_i}$. We also note that if the corresponding action is empty, it means $skip$.

For example, the \rc{t0} in Fig.~\ref{fig:srw_model} can be denoted as 
\begin{align*}
    & \tau [true]\langle skip \rangle \langle \{1.0/skip \to \\
    & \quad\lfloor steps=Update(steps, MaxSteps, x== 0) \rfloor Move\}\rangle
\end{align*}
where $\tau$ is an invisible event for an empty event in RoboChart models and $true$ is the default guard condition if a transition has no guard. 
The transition \rc{t1} can be denoted as 
\begin{align*}
    & \tau [x<MaxDist\land x>-MaxDist\land steps < MaxSteps]\\
    & \quad \langle skip\rangle \langle \{Pl/x=Minux(x, -MaxDist);\\
    & \quad \quad left \to \lfloor steps=\dots\rfloor Move\} + \\
    &\quad \{1-Pl/x=Plus(x, MaxDist);\\
    & \quad\quad right \to \lfloor steps=\dots\rfloor Move\} \rangle
\end{align*}
We note that the transition actions $a_{t}$ for both alternatives contain synchronisation events: $left$ or $right$.

After we have states in Markov models, we need to define transitions in Markov models in terms of transitions in RoboChart models. Figure~\ref{fig:robochart_trans_to_markov_trans} defines the Markov semantics for RoboChart transitions where we assume a RoboChart transition $t$ from a source node $s$ is enabled (that is, its guard condition is true and its trigger event is engaged).
\begin{figure*}\centering
  \subfloat[A simple transition without any action.]{\label{fig:trans_simple_no_action} 
  \begin{minipage}[b]{\linewidth}
\begin{align*}
    \begin{array}{c}
    t=e[g]\langle skip\rangle \langle \left\{1/skip \to \lfloor skip\rfloor s'\right\}\rangle 
    \\[\medskipamount] \cline{1-1} 
    \noalign{\smallskip}
    \mkvstate={\left(\setVars, \cdots, {\left(\setVars_{c_i}, \dots, {\left(\setVars_{m_j}, lk=0, \dots, {\left(pc=s, \_ \right)},\cdots\right)}, \cdots \right)}, \cdots, \right)} \to \\
    \mkvstate'={\left(\setVars, \cdots, {\left(\setVars_{c_i}, \dots, {\left(\setVars_{m_j}, lk=0, \dots, {\left({\cls pc=s'}, \_ \right)},\cdots\right)}, \cdots \right)}, \cdots, \right)}
    \end{array}
\end{align*}
  \end{minipage}
}
\\
  \subfloat[A simple transition entrying a state with an entry action.]{\label{fig:trans_simple_entry_action} 
  \begin{minipage}[b]{\linewidth}
\begin{align*}
    \begin{array}{c}
    t=e[g]\langle skip\rangle \langle \left\{1/skip \to \lfloor a_e\rfloor s'\right\}\rangle 
    \\[\medskipamount] \cline{1-1} 
    \noalign{\smallskip}
    \begin{array}[]{l}
    \mkvstate={\left(\setVars, \cdots, {\left(\setVars_{c_i}, \dots, {\left(\setVars_{m_j}, lk=0, \dots, {\left(pc=s, \_ \right)},\cdots\right)}, \cdots \right)}, \cdots, \right)} \to \\
    \mkvstate_1={\left(\setVars, \cdots, {\left(\setVars_{c_i}, \dots, {\left(\setVars_{m_j}, {\cls lk=t}, \dots, {\left({\cls pc=s'\_entering}, \_ \right)},\cdots\right)}, \cdots \right)}, \cdots, \right)} \to \\
    \mkvstate_2={\left({\cls \setVars[a_{e_1}]}, \cdots, {\left({\cls \setVars_{c_i}[a_{e_1}]}, \dots, {\left({\cls \setVars_{m_j}[a_{e_1}]}, {lk=t}, \dots, {\left({\cls pc=s'\_a_{e_1}}, \_ \right)},\cdots\right)}, \cdots \right)}, \cdots, \right)} \to \\
    \vdots \\
    \mkvstate'={\left({\cls \setVars[a_{e_1}]\dots[a_{e_p}]}, \cdots, {\left({\cls \setVars_{c_i}[a_{e_1}]\dots[a_{e_p}]}, \dots, {\left({\cls \setVars_{m_j}[a_{e_1}]\dots[a_{e_p}]}, {\cls lk=0}, \dots, {\left({\cls pc=s'}, \_ \right)},\cdots\right)}, \cdots \right)}, \cdots, \right)}
    \end{array}
    \end{array}
\end{align*}
  \end{minipage}
}
\\
  \subfloat[A transition with a transition action and entrying a state with an entry action.]{\label{fig:trans_simple_trans_action_entry_action} 
  \begin{minipage}[b]{\linewidth}
\begin{align*}
    \begin{array}{c}
    t=e[g]\langle skip\rangle \langle \left\{1/a_t \to \lfloor a_e\rfloor s'\right\}\rangle 
    \\[\medskipamount] \cline{1-1} 
    \noalign{\smallskip}
    \begin{array}[]{l}
    \mkvstate={\left(\setVars, \cdots, {\left(\setVars_{c_i}, \dots, {\left(\setVars_{m_j}, lk=0, \dots, {\left(pc=s, \_ \right)},\cdots\right)}, \cdots \right)}, \cdots, \right)} \to \\
    \mkvstate_1={\left(\setVars, \cdots, {\left(\setVars_{c_i}, \dots, {\left(\setVars_{m_j}, {\cls lk=t}, \dots, {\left({\cls pc=t\_action}, \_ \right)},\cdots\right)}, \cdots \right)}, \cdots, \right)} \to \\
    \mkvstate_2={\left({\cls \setVars[a_{t_1}]}, \cdots, {\left({\cls \setVars_{c_i}[a_{t_1}]}, \dots, {\left({\cls \setVars_{m_j}[a_{t_1}]}, {lk=t}, \dots, {\left({\cls pc=a_{t_1}}, \_ \right)},\cdots\right)}, \cdots \right)}, \cdots, \right)} \to \\
    \vdots \\
    \mkvstate_l={\left({\cls \setVars[a_{t_1}]\dots[a_{t_l}]}, \cdots, {\left({\cls \setVars_{c_i}[a_{t_1}]\dots[a_{t_l}]}, \dots, {\left({\cls \setVars_{m_j}[a_{t_1}]\dots[a_{t_l}]}, {lk=0}, \dots, {\left({\cls pc=s'\_entering}, \_ \right)},\cdots\right)}, \cdots \right)}, \cdots, \right)} \\
    \vdots \\
    \mkvstate'={\left({\cls \setVars[a_{t_1}]\dots[a_{t_l}][a_{e_1}]\dots[a_{e_p}]}, \cdots, {\left({\cls \setVars_{c_i}[a_{t_1}]\dots[a_{e_p}]}, \dots, {\left({\cls \setVars_{m_j}[a_{t_1}]\dots[a_{e_p}]}, {\cls lk=0}, \dots, {\left({\cls pc=s'}, \_ \right)},\cdots\right)}, \cdots \right)}, \cdots, \right)}
    \end{array}
    \end{array}
\end{align*}
  \end{minipage}
}
\\
\subfloat[A transition with multiple alternatives through a probabilistic junction.]{\label{fig:trans_prob_choice} 
   \begin{minipage}[b]{\linewidth}
 \begin{align*}
     \begin{array}{c}
     t=e[g]\langle skip \rangle \langle \left\{p_1/a_{t_1} \to \lfloor a_{e_1}\rfloor s_1\right\} + \dots + \left\{p_n/a_{t_n} \to \lfloor a_{e_n} \rfloor s_n\right\} \rangle 
     \\[\medskipamount] \cline{1-1} 
     \noalign{\smallskip}
    \begin{array}[]{l}
        \mkvstate={\left(\setVars, \cdots, {\left(\setVars_{c_i}, \dots, {\left(\setVars_{m_j}, lk=0, \dots, {\left(pc=s, \_\right)},\cdots\right)}, \cdots \right)}, \cdots, \right)} \to  \\
        \mkvstate_1={\left(\setVars, \cdots, {\left(\setVars_{c_i}, \dots, {\left(\setVars_{m_j}, {\cls lk=t}, \dots, {\left({\cls pc=pjunc}, \_\right)},\cdots\right)}, \cdots \right)}, \cdots, \right)} \to  \\
        \begin{cases}
             p_1: \left\{ 
             \begin{array}[]{l}
                 \mkvstate_2'={\left({\setVars}, \cdots, {\left({\setVars_{c_i}}, \dots, {\left({\setVars_{m_j}}, {lk=t}, \dots, {\left({\cls pc=t\_action}, \_\right)},\cdots\right)}, \cdots \right)}, \cdots, \right)} \\
                 \vdots \\
             \end{array}\right. &  \\
            \vdots & \\ 
             p_n: \left\{ 
             \begin{array}[]{l}
                 \mkvstate_2'={\left({ \setVars}, \cdots, {\left({\setVars_{c_i}}, \dots, {\left({\setVars_{m_i}}, {lk=t}, \dots, {\left({\cls pc=t\_action}, \_\right)},\cdots\right)}, \cdots \right)}, \cdots, \right)} \\
                 \vdots \\
             \end{array}\right. & \\
         \end{cases}
     \end{array}
     \end{array}
 \end{align*}
   \end{minipage}
 }
 \\
  \subfloat[A transition exiting from a state with an exit action.]{\label{fig:trans_simple_exit_action} 
  \begin{minipage}[b]{\linewidth}
\begin{align*}
    \begin{array}{c}
    t=e[g]\langle a_x \rangle \langle \left\{1/skip \to \lfloor skip\rfloor s'\right\}\rangle 
    \\[\medskipamount] \cline{1-1} 
    \noalign{\smallskip}
    \begin{array}[]{l}
        \mkvstate={\left(\setVars, \cdots, {\left(\setVars_{c_i}, \dots, {\left(\setVars_{m_j}, lk=0, \dots, {\left(pc=s, exit=\text{NONE}\right)},\cdots\right)}, \cdots \right)}, \cdots, \right)} \to \\
        \mkvstate_1={\left(\setVars, \cdots, {\left(\setVars_{c_i}, \dots, {\left(\setVars_{m_j}, {\cls lk=t}, \dots, {\left(pc=s, {\cls exit=\text{Sub\_ACT}}\right)},\cdots\right)}, \cdots \right)}, \cdots, \right)} \to \\
    \mkvstate_2={\left({\cls \setVars[a_{x_1}]}, \cdots, {\left({\cls \setVars_{c_i}[a_{x_1}]}, \dots, {\left({\cls \setVars_{m_j}[a_{x_1}]}, {lk=t}, \dots, {\left({\cls pc=a_{x_1}}, {exit=\text{Sub\_ACT}}\right)},\cdots\right)}, \cdots \right)}, \cdots, \right)} \to \\
    \vdots \\
    \mkvstate_p={\left({\cls \setVars[a_{x_1}]\dots[a_{x_p}]}, \cdots, {\left({\cls \setVars_{c_i}[a_{x_1}]\dots[a_{x_p}]}, \dots, {\left(
    \begin{array}[]{l}
        {\cls \setVars_{m_j}[a_{x_1}]\dots[a_{x_p}]}, {lk=t}, \dots, \\
        {\left({\cls pc=a_{x_p}}, {\cls exit=\text{Sub\_EXITED}}\right)},\cdots
    \end{array}
    \right)}, \cdots \right)}, \cdots, \right)} \to \\
    \mkvstate_{p+1}={\left({\setVars[a_{x_1}]\dots[a_{x_p}]}, \cdots, {\left({\setVars_{c_i}[a_{x_1}]\dots[a_{x_p}]}, \dots, {\left(
    \begin{array}[]{l}
        {\setVars_{m_j}[a_{x_1}]\dots[a_{x_p}]}, {lk=t}, \dots, \\
        {\left({\cls pc=s'\_entering}, {\cls exit=\text{NONE}}\right)},\cdots
    \end{array}
    \right)}, \cdots \right)}, \cdots, \right)} \to \\
    \vdots \\
    \mkvstate'={\left({\setVars[a_{x_1}]\dots[a_{x_p}]}, \cdots, {\left({\setVars_{c_i}[a_{x_1}]\dots[a_{x_p}]}, \dots, {\left(
    \begin{array}[]{l}
        {\setVars_{m_j}[a_{x_1}]\dots[a_{x_p}]}, {\cls lk=0}, \dots, \\
        {\left({\cls pc=s'}, {exit=\text{NONE}}\right)},\cdots
    \end{array}
    \right)}, \cdots \right)}, \cdots, \right)} \to \\
    \end{array}
    \end{array}
\end{align*}
  \end{minipage}
}
\\
\caption{The semantics of RoboChart transitions in Markov models.}\label{fig:robochart_trans_to_markov_trans}
\end{figure*}

Figure~\ref{fig:trans_simple_no_action} shows a simple RoboChart transition $t$ (above the line) which has a trigger $e$ and a guard $g$, but no transition action. The target node of $t$ is $s'$. There is no exit action in $s$ and no entry action in $s'$. The corresponding Markov transition from state $\mkvstate$ to $\mkvstate'$ is shown below the line. The only changed variable in $\mkvstate'$ is $pc$ for the composite state that contains the $t$ (also $s$ and $s'$). For example, the transition \rc{t6} in Fig.~\ref{fig:srw_model} is a such simple transition. Its corresponding transition in Markov models is $\left(x, steps, lk, pc=Move\right) \to \left(x, steps, lk, pc=Stuck\right)$, which denotes the counter ($pc$) moves from \rc{Move} to \rc{Stuck}.

In Fig.~\ref{fig:trans_simple_entry_action}, we consider $s'$ has an entry action $a_e$ which is sequential composition of $p$ actions (so $a_e=a_{e_1};\dots a_{e_p}$). The first transition (from $\mkvstate$ to $\mkvstate_1$) in Markov models sets $lk$ to $t$ (that is, the machine is locked in taking $t$ so other transitions from the same machine cannot take) from 0 (that is, the lock if free), and $pc$ to $s'\_entering$ (an intermediate state for $s'$ to denote a transition is about to enter $s'$). Then the next transition (from $\mkvstate_1$ to $\mkvstate_2$) corresponds to the execution of the action $a_{e_1}$ where $pc$ is set to another intermediate state $s'\_a_{e_1}$. At the same time, the shared variables $\setVars$, the variables $\setVars_{c_i}$ and $\setVars_{m_j}$ declared in the controller and machine that contain $t$, could be updated by $a_{e_1}$, denoted as $\setVars[a_{e_1}]$. The subsequent transitions corresponds to the executions of other actions in $a_e$. The final transition to $\mkvstate'$ sets $lk$ back to 0 (so other corresponding RoboChart transitions can take), $pc$ to $s'$ (so $s'$ is entered), and updates the corresponding variables by the last action $a_{e_p}$. 
The transition \rc{t0} in Fig.~\ref{fig:srw_model} is a such simple transition. Its corresponding transitions in Markov models are 
\begin{align*}
    & \left(x, steps, lk=0, pc=i0\right) \to \\
    & \left(x, steps, lk=t0, pc=Move\_entering\right) \to\\
    & \left(x, steps=Update(\dots), lk=0, pc=Move\right)
\end{align*}
Finally, the \rc{Move} state is entered with \rc{steps} updated to the result of the application of the \rc{Update} function to the arguments which is omitted here.

In Fig.~\ref{fig:trans_simple_trans_action_entry_action}, we further consider a transition with a transition action $a_t$. Now $\mkvstate_1$ corresponds to the start of the execution of $a_t$, and $\mkvstate_l$ corresponds to the completion of $a_t$ and the start of the entry action $a_e$.
The transition \rc{t4} in Fig.~\ref{fig:srw_model} is a such transition. Its corresponding transitions in Markov models are 
\begin{align*}
    & \left(x, steps, lk=0, pc=Move\right) \to  \\
    & \left(x, steps, lk=t4, pc=t4\_act\right) \to \\
    & \left(Minus(x, -MaxDist), steps, t4, t4\_act\_1\right) \to \\
    & \left(Minus(x, -MaxDist), steps, t4, Move\_entering\right) \to \\
    & \left(Minus(x, -MaxDist), Update(\dots), 0, Move\right)
\end{align*}
Finally, both $x$ and $steps$ are updated. 

RoboChart transitions can be probabilistic, such as the transitions \rc{t1} and its two probabilistic alternatives \rc{t2} and \rc{t3} in Fig.~\ref{fig:srw_model}. Its corresponding Markov transitions are shown in Fig.~\ref{fig:trans_prob_choice} where $t$ is a general transition with $n$ alternatives, of which the $i$th alternative with probability $p_i$, and additionally $\sum_{i=1}^{n} p_i = 1$. The first Markov transition (from $\mkvstate$ to $\mkvstate_1$) sets $lk$ to $t$ and updates $pc$ to an intermediate state $pjunc$, representing the probabilistic junction. After that, the transition is a probabilistic choice with corresponding probabilities. For the transition \rc{t1} and alternatives \rc{t2} and \rc{t3} in Fig.~\ref{fig:srw_model}, the corresponding Markov transitions are shown below where $p0$ is the name of the probabilistic junction.
\begin{align*}
    & \left(x, steps, lk=0, pc=Move\right) \to  \\
    & \left(x, steps, lk=t1, pc=p0\right) \to \\
    & \left\{
    \begin{array}[]{l}
    Pl: \\
    \left\{
        \begin{array}[]{l}
            \left(x, steps, lk=t1, pc=t3\_act\_1\right) \to \\
            \left(Minus(x, -MaxDist), steps, t1, t3\_act\_2\right) \to \\
            \left(Minus(x, -MaxDist), steps, t1, Move\_entering\right) \to \\
            \left(Minus(x, -MaxDist), Update(\dots), 0, Move\right)
        \end{array}
        \right.
        \\
    (1-Pl): \\
    \left\{
        \begin{array}[]{l}
            \left(x, steps, lk=t1, pc=t2\_act\_1\right) \to \\
            \left(Plus(x, MaxDist), steps, t1, t2\_act\_2\right) \to \\
            \left(Plus(x, MaxDist), steps, t1, Move\_entering\right) \to \\
            \left(Plus(x, MaxDist), Update(\dots), 0, Move\right)
        \end{array}
        \right.
    \end{array}
    \right. 
\end{align*}

In Fig.~\ref{fig:trans_simple_exit_action}, we consider a transition whose source state $s$ has an exit action $a_x$. Now $\mkvstate_1$ corresponds to the start of the exit from $s$ (that is, $exit=\text{Sub\_ACT}$), and $\mkvstate_2$ and $\mkvstate_p$ correspond to the start and completion of the execution of $a_x$. After the completion of $a_x$, $s$ is exited (that is, $exit=\text{Sub\_EXITED}$). The next state $\mkvstate_{p+1}$ marks the start of entering $s'$ with the completion of exiting from $s$ (that is, $exit=\text{NONE}$).

RoboChart models contain a high degree of nondeterminism which could arise from multiple transitions from a source node with the same trigger and simultaneously enabled guard conditions, or from different hierarchical states (e.g. parent state and its substates), or from different state machines or controllers. Figure~\ref{fig:trans_nondeter} shows if $n$ transitions are enabled, their corresponding Markov transitions are nondeterministically (or uniformly) chosen for MDP (or DTMC) models. Another nondeterminism is introduced because RoboChart transitions are non-atomic (due to their actions). For example, when a transition $t_1$ is taken in a machine $m_1$ but has not yet entered its target state, another transition $t_2$ is also taken in another machine $m_2$. Nondeterminism arises from the situation: of which transition the action should be chosen to execute next: either $t_1$ or $t_2$?
\begin{figure*}\centering
\subfloat[Nondeterminism (multiple enabled transitions in a state machine or between state machines)]{\label{fig:trans_nondeter} 
  \begin{minipage}[b]{\linewidth}
\begin{align*}
    \begin{array}{c}
        \begin{array}[]{l}
            t_1=e[g_1]\langle a_x\rangle \langle \dots \rangle \qquad 
            \dots \qquad 
            t_n=e[g_n]\langle a_x\rangle \langle \dots \rangle 
        \end{array}
    \\[\medskipamount] \cline{1-1} 
    \noalign{\smallskip}
    \begin{array}[]{l}
        \mkvstate={\left(\setVars, \cdots, {\left(\setVars_{c_i}, \dots, {\left(\setVars_{m_j}, lk=0, \dots, {\left(pc=s, \_\right)},\cdots\right)}, \cdots \right)}, \cdots, \right)} \to  
        \left\{
        \begin{array}[]{l}
            \mkvstate_{11} \to \dots \to \mkvstate_1' \\
            \vdots \\
            \mkvstate_{n1}  \to \dots \to \mkvstate_n' \\
        \end{array}
        \right\} \text{nondeterministic choice}
    \end{array}
    \end{array}
\end{align*}
  \end{minipage}
}
\\
\subfloat[Synchronisation (two transitions from two different machines)]{\label{fig:sync} 
  \begin{minipage}[b]{\linewidth}
\begin{align*}
    \begin{array}{c}
        \begin{array}[]{l}
            t_1=e[g_1]\langle a_{x_1}\rangle \langle \dots \rangle \qquad 
            t_2=e[g_2]\langle a_{x_2}\rangle \langle \dots \rangle 
        \end{array}
    \\[\medskipamount] \cline{1-1} 
    \noalign{\smallskip}
    \begin{array}[]{l}
        \mkvstate={\left(\setVars, \cdots, {\left(\setVars_{c_i}, \dots, {\left(\setVars_{m_j}, lk=0, \dots, {\left(pc=s, \_\right)},\cdots\right)}, \cdots, {\left(\setVars_{m_k}, lk=0, \dots, {\left(pc=s, \_\right)},\cdots\right)}, \cdots \right)}, \cdots, \right)} \to  \\
        \mkvstate_1={\left(\setVars, \cdots, {\left(\setVars_{c_i}, \dots, 
        \left\{
            \begin{array}[]{l}
                {\left(\setVars_{m_j}, {\cls lk=t_1}, \dots, {\left({\cls pc=s_1}, {\cls exit=\text{Sub\_ACT}}\right)},\cdots\right)}, \cdots, \\
                {\left(\setVars_{m_k}, {\cls lk=t_2}, \dots, {\left({\cls pc=s_2}, {\cls exit=\text{Sub\_ACT}}\right)},\cdots\right)}, \cdots 
            \end{array}
        \right\}
            \right)}, \cdots, \right)} \to  \\
        \vdots \\
    \end{array}
    \end{array}
\end{align*}
  \end{minipage}
}
\\
\subfloat[Communication (two transitions from two different machines)]{\label{fig:comms} 
  \begin{minipage}[b]{\linewidth}
\begin{align*}
    \begin{array}{c}
        \begin{array}[]{l}
            t_1=e!v[g_1]\langle a_{x_1}\rangle \langle \dots \rangle \qquad
            t_2=e?x[g_2]\langle a_{x_2}\rangle \langle \dots \rangle 
        \end{array}
    \\[\medskipamount] \cline{1-1} 
    \noalign{\smallskip}
    \begin{array}[]{l}
        \mkvstate={\left(\setVars, \cdots, {\left(\setVars_{c_i}, \dots, {\left(\setVars_{m_j}, lk=0, \dots, {\left(pc=s, \_\right)},\cdots\right)}, \cdots, {\left(\setVars_{m_k}, lk=0, \dots, {\left(pc=s, \_\right)},\cdots\right)}, \cdots \right)}, \cdots, \right)} \to  \\
        \mkvstate_1={\left(\setVars, \cdots, {\left(\setVars_{c_i}, \dots, 
        \left\{
            \begin{array}[]{l}
                {\left(\setVars_{m_j}, {\cls lk=t_1}, \dots, {\left({\cls pc=s_1}, {\cls exit=\text{Sub\_ACT}}\right)},\cdots\right)}, \cdots, \\
                {\left({\cls \setVars_{m_k}[x=v]}, {\cls lk=t_2}, \dots, {\left({\cls pc=s_2}, {\cls exit=\text{Sub\_ACT}}\right)},\cdots\right)}, \cdots 
            \end{array}
        \right\}
            \right)}, \cdots, \right)} \to  \\
        \vdots \\
    \end{array}
    \end{array}
\end{align*}
  \end{minipage}
}
\caption{The semantics of RoboChart transitions with communication in Markov models.}\label{fig:robochart_trans_to_markov_trans_comms}
\end{figure*}

RoboChart allows communication between two connection nodes (machines, controllers, or robotic platforms) through connections. Figure~\ref{fig:sync} shows two transitions $t_1$ and $t_2$ from different state machines $m_j$ and $m_k$, which synchronise over a same trigger event $e$ (excluding $\tau$). Its first corresponding Markov transition (to $u_1$) sets $lk$ of $m_j$ to $t_1$ and of $m_k$ to $t_12$. In Fig.~\ref{fig:comms}, two transitions $t_1$ and $t_2$ from different state machines $m_j$ and $m_k$ communicate over a same trigger event $e$ for input $e?x$ and output $e!v$. Its first corresponding Markov transition (to $u_1$) is similar to that of the synchronisation except the variable $x$ in $\setVars_{m_k}$ of $m_k$ is updated to $v$ from $m_j$.

\subsection{PCTL* for specifying RoboChart properties}
\label{sec:markov:pctl}
After the interpretation of RoboChart semantics in Markov models, we can use PCTL*~\cite{Aziz1995,Bianco1995,Baier1998} to specify properties for RoboChart models. We show the syntax of PCTL* below where $\bowtie \in \{<, >, \leq, \geq\}$, $k \in \mathbb{N}$, and $p \in [0..1]$.

\begin{align*}
    & \phi~::=~true ~|~ a  ~|~ \lnot \phi  ~|~ \phi_1 \land \phi_2 ~|~ \pctlOp{A} \psi ~|~ \pctlOp{E} \psi ~|~ \pctlOp{P}_{\bowtie p}(\psi)  \tag*{(state formulas)}\\
    & \psi~::=~\phi ~|~ \lnot \psi  ~|~ \psi_1 \land \psi_2 ~|~  \pctlOp{X} \psi ~|~  \psi_1 \,\pctlOp{U}\,\psi_2 ~|~  \psi_1\,\pctlOp{U}^{\leq k}\,\psi_2 \tag*{(path formulas)} 
\end{align*}

From these path formulas, we can derive others $\pctlOp{F}$ (eventually), $\pctlOp{G}$ (always),  $\pctlOp{W}$ (weak until),  $\pctlOp{R}$ (release), and their bounded variants. 
\begin{align*}
    & \pctlOp{F} \psi = true~\pctlOp{U} \psi \quad 
     \pctlOp{F}^{\leq k} \psi = true~\pctlOp{U}^{\leq k} \psi \\
     & \pctlOp{G} \psi = \lnot \left(\pctlOp{F} \lnot \psi\right)\quad
     \pctlOp{G}^{\leq k} \psi = \lnot \left(\pctlOp{F}^{\leq k} \lnot \psi\right)  \\
    & \psi_1\,\pctlOp{W}\,\psi_2 = \left(\psi_1\,\pctlOp{U}\,\psi_2 \right) \lor \left(\pctlOp{G} \psi_1\right) \\
    & \psi_1\,\pctlOp{W}^{\leq k}\,\psi_2 = \left(\psi_1\,\pctlOp{U}^{\leq k}\,\psi_2 \right) \lor \left(\pctlOp{G}^{\leq k} \psi_1\right) \\ 
    & \psi_1\,\pctlOp{R}\,\psi_2 = \lnot \left(\lnot \psi_1\,\pctlOp{U}^{\leq k}\,\lnot\psi_2 \right)
\end{align*}

For a MDP $\left(S, s_{init}, Act, Steps, L\right)$, we define a path as 
\begin{align*}
& \pi=s_0(a_0,\mu_0)s_1(a_1,\mu_1)\dots s_i(a_i,\mu_i)\dots
\end{align*}
such that $(a_i,\mu_i) \in Steps(s_i)$ and $\mu_i(s_{i+1}) > 0$ 
where $a_i$ is the action used to resolve nondeterminism at state $s_i$ and $\mu_i$ is the probability distribution associated with $a_i$ in state $s_i$. We define $\xi(\pi)=s_0s_1\dots s_i\dots$ to extract a sequence of states from a path, the ith element in a path as $\pi_i = s_i(a_i,\mu_i)$, and the length of a path as $|\pi|=|\xi(\pi)|$ (that is, the number of states in a path). We further define notations to extract a state, an action, or a distribution from $\pi_i$: $\pi_i^s = s_i$, $\pi_i^a = a_i$, and $\pi_i^\mu = \mu_i$. We use $Path(s)$ to denote all (finite or infinite) paths from state $s$ in the MDP model.

An adversary $\mathcal{\sigma}$ of a MDP is a function mapping every finite path $\pi$, denoted as $s_0(a_0,\mu_0)s_1(a_1,\mu_1)\dots s_n$, to an element of $Steps(\pi_n^s)$. Intuitively, it resolves nondeterministic choice in the last state $s_n$ by choosing one of its action and distribution pair $\left(a_n, \mu_n\right)$ which belongs to $Steps(\pi_n^s)$. We note that $\sigma$ only resolves nondeterministic choice not probabilistic choice by $\mu_n$. We use $\mathcal{A}$ to denote the set of all adversaries and $Path^\sigma(s)$ for paths from $s$ where nondeterminism is resolved by $\sigma$. 

We use $s \models \phi$ to denote the state formula $\phi$ is satisfied in state $s$, and $\pi \models
\psi$ to denote the path formula $\psi$ is satisfied in a path $\pi$. 
We show the semantics of state formulas below. 
\begin{align*}
    \begin{array}[]{lcl}
     s \models true & \iff & true \\
     s \models a & \iff & a \in L(s) \\
     s \models \lnot \phi & \iff & \lnot \left(s \models \phi\right) \\
     s \models \phi_1 \land \phi_2 & \iff & \left(s \models \phi_1\right) \land \left(s \models \phi_2\right) \\
     s \models \pctlOp{A} \psi & \iff & \forall \pi \in Path(s) \bullet \left(\pi \models \psi\right) \\
     s \models \pctlOp{E} \psi & \iff & \exists \pi \in Path(s) \bullet \left(\pi \models \psi\right) \\
     s \models \pctlOp{P}_{\bowtie p}(\psi) & \iff & \forall \sigma: \mathcal{A} \bullet Prob^\sigma(s, \psi) \bowtie p \\
    \end{array}
\end{align*}
In the semantics of the $\pctlOp{P}$ operator, 
\begin{align*}
    Prob^\sigma(s, \psi)=Pr_s\left\{\pi \in Path^\sigma(s)~\mid~\pi \models \psi \right\}
\end{align*}
where $Pr_s$ is the probability measure in a probability space over paths from $s$ ($Path(s)$). This gives the probability of the paths $\pi$ from $s$ that are resolved by $\sigma$ and also satisfy $\psi$. We also define the minimum and maximum probabilities as the greatest lower bound ($\pctlOp{inf}$) and the least upper bound ($\pctlOp{sup}$) of probabilities for all adversaries. 
\begin{align*}
    & \pctlOp{P_{min}}(s, \psi) = \pctlOp{inf}_{\sigma \in \mathcal{A}} Prob^\sigma(s, \psi) \\
    & \pctlOp{P_{max}}(s, \psi) = \pctlOp{sup}_{\sigma \in \mathcal{A}} Prob^\sigma(s, \psi) 
\end{align*}

The semantics of path formulas is shown as follows.
\begin{align*}
     & \pi \models \psi  \iff \pi_0^s \models \psi \\
     & \pi \models \lnot \psi  \iff  \lnot \left(\pi \models \psi\right) \\
     & \pi \models \psi_1 \land \psi_2  \iff  \left(s \models \psi_1\right) \land \left(s \models \psi_2\right) \\
     & \pi \models \pctlOp{X}\psi  \iff  \pi_1^s \models \psi \\
     & \pi \models \psi_1 \pctlOp{U}\psi_2  \iff  \\
     &\qquad\exists i \geq 0 \mid i \leq |\psi| \bullet \left(
     \begin{array}[]{l}
        \forall j \in [0..i-1] \bullet \pi_j^s \models \psi_1 \\
        \land \pi_i^s \models \psi_2 
     \end{array} \right) \\
     & \pi \models \psi_1 \pctlOp{U}^{\leq k} \psi_2  \iff  \\
     &\qquad \exists i \geq 0 \mid i \leq min(|\pi|, k) \bullet \left(
     \begin{array}[]{l}
        \forall j \in [0..i-1] \bullet \pi_j^s \models \psi_1 \\
        \land \pi_i^s \models \psi_2
    \end{array}\right)
\end{align*}

Using PCTL*, we can express both qualitative properties such as safety, liveness, and fairness properties, and quantitative properties.
\begin{example}[Interesting properties for the simple random walk]
    \label{ex:srw_pctl_properties}
    As discussed previously, the initial state $\mkvstate_{init}$ of a MDP for the simple random walk in Sect.~\ref{sec:robochart} is 
    \begin{align*}
        & \left(x=0,steps=0,lk=0,pc=i0\right)
    \end{align*}
    and 
    \begin{align*}
      &L\left(\mkvstate_{init}\right) = \left\{(x=0),(steps=0),(lk=0),(pc=i0)\right\}
    \end{align*}
    We consider qualitative properties below. 
    \begin{align*}
        & \mkvstate_{init} \models (pc=i0)  \tag*{(true)} \\
        & \mkvstate_{init} \models \pctlOp{X}(pc=Move)  \tag*{(false due to intermediate state $Move\_entering$)} \\
        & \mkvstate_{init} \models \pctlOp{A}\left(\pctlOp{F} \left(pc=Stuck\right)\right)  \tag*{(true or false)}
    \end{align*}
    The third formula could be true or false depending on the instantiation of the constant variable \rc{MaxSteps}. If it is equal to 0 (though it is not what we aim for), the property is true because the only transition enabled in state \rc{Move} is \rc{t6}, which takes the state machine to state \rc{Stuck}. Otherwise, it is possible that the state machine can repeatedly take one of the transitions \rc{t1}, \rc{t4}, or \rc{t5} without letting \rc{steps} reach \rc{MaxSteps} (so not enable \rc{t6}). Therefore, the property is false. For the properties below, we only consider the constants \rc{MaxSteps}, \rc{MaxDist},  and \rc{Pl} are all positive.
    \begin{align*}
        & \mkvstate_{init} \models \pctlOp{E}\left(\pctlOp{G} \left(\lnot pc=Stuck\right)\right)  \tag*{(true)} \\
        & \mkvstate_{init} \models \pctlOp{A}\left(\pctlOp{F} \pctlOp{G} \left(pc=Stuck\right)\right)  \tag*{(false)} \\
        & \mkvstate_{init} \models \pctlOp{E}\left(\pctlOp{F} \pctlOp{G} \left(pc=Stuck\right)\right)  \tag*{(true)} \\
        & \mkvstate_{init} \models \pctlOp{A}\left(\pctlOp{G} \pctlOp{F} \left(pc=Move\right)\right)  \tag*{(fairness, false)} \\
        & \mkvstate_{init} \models \pctlOp{A}\left(\pctlOp{G} \pctlOp{F} \left(pc=Move\right) \implies \pctlOp{G} \pctlOp{F} \left(pc=p0\right)\right)  \tag*{(strong fairness, true)} \\
        & \mkvstate_{init} \models \pctlOp{A}\left(\pctlOp{F} \pctlOp{G} \left(pc=Stuck\right) \implies \pctlOp{G} \pctlOp{F} \left(steps=MaxSteps\right)\right)  \tag*{(weak fairness, true)} \\
        & \mkvstate_{init} \models \pctlOp{A}\left(\pctlOp{G} \left(\left(steps=MaxSteps\right) \implies \pctlOp{F} (pc=Stuck)\right)\right)  \tag*{(liveness, true)} \\
        & \mkvstate_{init} \models \pctlOp{A}\left(\pctlOp{G} \left(x \leq MaxDist \land x \geq -MaxDist\right)\right)  \tag*{(safety and bounded, true)}
    \end{align*}

    We note in the last formula shown above, the expressions like $x \leq MaxDist \land x \geq -MaxDist$ are not an AP of the MDP, and so the formula is not a PCTL* formula. It is a short (or predicate) form of the set expression 
    \[\left\{v: \mathbb{N} | \left(v \leq MaxDist \land v \geq -MaxDist\right) \bullet (x=v)\right\}\]
The formula can be rewritten to a PCTL* formula below.
    \begin{align*}
        & \mkvstate_{init} \models \pctlOp{A}\left(\pctlOp{G} \left(
        \begin{array}[]{l}
            x=-MaxDist \lor \dots \lor x=-1 \lor x=0 \\
            \lor x=1 \lor \dots \lor x=MaxDist 
        \end{array}\right)\right)  \tag*{(Equivalent)}
    \end{align*}
However, it is not convenient to write a property like this way to list all possible values. The predicate form is more compact.

Using the $\pctlOp{P}$ operator, we can also specify quantitative properties.  
    \begin{align*}
        & \mkvstate_{init} \models \pctlOp{P}_{\leq 0.1}\left(\pctlOp{F} \left(pc=Stuck \right)\right)  \tag*{(The probability of being in \rc{Stuck})} \\
        & \mkvstate_{init} \models \pctlOp{P}_{\leq 0.1}\left(\pctlOp{F}^{\leq 10} \left(pc=Stuck \right)\right)  \tag*{(The probability being \rc{Stuck} after 10 units of time)} \\
        & \mkvstate_{init} \models \pctlOp{P_{min}}\left(\mkvstate_{init}, \pctlOp{G} \pctlOp{F} (pc=Move) \right) \geq 0.9 \tag*{(The minimum probability of not getting \rc{Stuck})}
    \end{align*}
    \qed
\end{example}

\subsection{Motivations of RoboCertProb}
\label{sec:markov:motivation}
RoboChart aims for roboticists to write models for robot software controllers and properties for verification. There are a number of challenges and problems in using PCTL* to write properties for RoboChart models directly. The major challenge is the requirement of knowledge of RoboChart semantics in Markov models in order for them to use appropriate formulas to capture their properties. The second problem is the use of non-RoboChart elements in Markov semantics, such as $pc$, $lk$, and $exit$ variables that are introduced only for Markov semantics. They are invisible in RoboChart models, which makes it difficult and non-intuitive to specify properties in terms of non-RoboChart elements. Thirdly, RoboChart models are loosely specified using constants (or the parameters of models), and unspecified functions and operations. Properties need to be specified in the context of a particular instance of the model. However, PCTL* is based on the (implicit) assumptions of an instantiation we made for RoboChart models. 
Another problem is to choose the right variables in Markov models for specifying properties due to RoboChart's component model for reuse (so the mapping from RoboChart elements to Markov models is not one-to-one). We have also seen the convenience of using the predicate form for specifying APs, and so a rich syntax for expressions does simplify specifications. It is worth mentioning that one RoboChart transition may correspond to multiple Markov transitions, as shown in Figs.~\ref{fig:robochart_trans_to_markov_trans} and \ref{fig:robochart_trans_to_markov_trans_comms}, which makes the bounded variants of $\pctlOp{U}$, $\pctlOp{F}$, and $\pctlOp{G}$ (where the bounded number in these operators means the number of transitions, or units of discrete time, taken) not very useful because the discrete time in Markov models has no direct meaning in their RoboChart models. For this reason, we should not use bounded variants for specifications (though it might be useful for simulation or testing).

These practical problems motivate us to design a property language suitable for roboticists to use. It should allow them  
\begin{enumerate*}[label={(\arabic*)}]
    \item to specify the context (or instantiation) of a RoboChart model for properties to be verified,  
    \item using only RoboChart elements, 
    \item with a controlled language syntax and rich expressions to facilitate usage, 
    \item by fully automated tool support. 
\end{enumerate*}

Our first option to consider is the PRISM's property specification (PrPS).\footnote{\url{www.prismmodelchecker.org/manual/PropertySpecification}} This is due to a fact that our semantics for RoboChart models is defined through transformation rules~\cite{Ye2022} from RoboChart to PRISM, instead of directly given in Markov models, as defined in this section. This transformation has been implemented in RoboTool. The use of PrPS for specification could be easily applied to PRISM. It addresses the rich expressions problem in PCTL*, but other challenges remain due to the differences between RoboChart and PRISM, particularly at the abstraction level~\cite{Ye2022}.
The language RoboCertProb is designed to address these problems.

\section{Probabilistic property language: RoboCertProb}
\label{sec:ppl_syntax}

This section presents the probabilistic property language (RoboCertProb) in RoboCert, a language designed to facilitate the verification of RoboChart models.


RoboCertProb specifies properties of RoboChart models in terms of RoboChart elements instead of those in the generated semantics models, so users focus on modelling and verification using RoboChart and RoboTool. The RoboCertProb grammar presented in the sections from Sect.~\ref{ssec:ppl_syntax_prob_statements} will reflect these.

RoboChart has a component model, and its semantics is compositional, while the PRISM language has a flat structure and is not compositional. Because of this difference, RoboChart's structure is flattened when transformed to PRISM, and all RoboChart elements must be uniquely identified through potential references~\cite{Ye2022}.
Furthermore, events in RoboChart can be bidirectionally used for both input and output. Their corresponding actions in PRISM for each bidirectional event are two: one for input and one for output. For this reason, we also need a way to specify RoboChart events with directions. However, this unique reference mechanism is not directly provided by RoboChart and RoboTool. We implement a new reference mechanism called fully qualified names to RoboChart elements, as described in Sect.~\ref{ssec:ppl_syntax_fqfnelem}.

RoboChart models are loosely specified through constant variables, functions and operation definitions. They must be instantiated in verification models, described in Sects.~\ref{ssec:ppl_syntax_constconfig} and \ref{ssec:ppl_syntax_func_op_defs}.

RoboChart models reactive robotic systems, while PRISM has a closed-world assumption. To specify the RoboChart model's behaviour when subjected to a particular input, we need to set up a specific environmental input and check its expected behaviour for generated PRISM models. This is discussed in Sect.~\ref{ssec:ppl_syntax_environment}.

Other aspects of RoboCertProb correspond to those of PrPS:
\begin{enumerate*}[label={(\alph*)}]
\item labels and formulas are for reuse,
\item quantitative properties are specified using probabilistic (P) and reward (R) operators,
\item qualitative properties are expressed in path quantifiers: for all paths (A) and some paths (E) as in CTL*, and
\item statistical model checking (SMC) or simulation is for the models with larger state space.
\end{enumerate*}

The semantics of RoboCertProb is based on PCTL*, as presented in Sect.~\ref{sec:markov:pctl}, and interpreted in the discrete part of PrPS. Semantically, RoboCertProb can be seen as a subset of PrPS.

Next, we present the grammar of RoboCertProb using a combination of Extend Backus Naur Form (EBNF) and Xtext,\footnote{\url{www.eclipse.org/Xtext/documentation/301_grammarlanguage.html}} an Eclipse-based open-source framework for developing domain-specific languages.


\subsection{Probabilistic statements}

\label{ssec:ppl_syntax_prob_statements}

The syntax of RoboCertProb is defined by the nonterminal \lstinbnf{ProbStatements}, which contains a sequence of probabilistic statements \lstinbnf{ProbStatement} shown below.

\begin{lstlisting}[language=AssertBNF,]
ProbStatements ::= /*(package QFN)? */ProbStatement*/*
QFN            ::= ID ('::' ID)* */
ProbStatement  ::= /*Import | */Constant | Constants | Label | Formula | Rewards | Definitions | pModules | ProbProperty
\end{lstlisting}

\lstinbnf{ProbStatement} has various alternatives:
\begin{enumerate*}[label={(\alph*)}]
\item \lstinbnf{Constant} for declaring constant variables used in properties;
\item \lstinbnf{Constants} for configuring constant variables;
\item \lstinbnf{Label} for defining labels to identify a set of states that are of particular interest;
\item \lstinbnf{Formula} for defining formulas, a shorthand to expressions for reuse;
\item \lstinbnf{Rewards} for defining rewards or costs that are associated with states and transitions;
\item \lstinbnf{Definitions} for defining loose functions and operations in RoboChart models; 
\item \lstinbnf{pModules} for corresponding PRISM modules to specify the environment of RoboChart models; and 
\item \lstinbnf{ProbProperty} to specify probabilistic properties. 
\end{enumerate*}

The syntax of \lstinbnf{Constant}, \lstinbnf{Label}, \lstinbnf{Formula}, and \lstinbnf{Rewards} are straightforward and omitted here for simplicity. We refer to the RoboChart reference manual~\cite[Sect.~6.2]{A.Miyazawa2020} for a complete account of RoboCertProb.

Before presenting the details of RoboCertProb, we first define how RoboChart model elements are referred to in properties.

%
%


\subsection{References to RoboChart model elements}
\label{ssec:ppl_syntax_fqfnelem}


RoboCertProb specifies properties in terms of RoboChart models where model elements are specified using fully qualified names \lstinbnf{FQFNElem}, with multiple segments, shown below.
\begin{lstlisting}[language=AssertBNF,]
FQFNElem ::= ([RCPackage] | [NamedElement]) ('::' [NamedElement])*
\end{lstlisting}

Xtext's syntax defines a cross reference by the text inside the square brackets. The text refers not to another rule but to a class in the RoboChart's metamodel. \lstinbnf{RCPackage} is the root class of a RoboChart model, and \lstinbnf{NamedElement} is the parent class of all classes in RoboChart that have names, such as modules, states, and transitions.

It is important to note that the rule is flexible, but not all qualified names satisfying this rule are valid RoboChart model elements. For example, \lstinbnf{SRWCtrl::SRWMod} is not valid because the module \rc{SRWMod} is not a sub-element of \rc{SRWCtrl}. These cases are ruled out by the well-formedness conditions defined below, and these conditions are implemented in tools for validation, which will be discussed in Sect.~\ref{sec:tool}.  \smallskip
\begin{enumerate}[label=WFREF-{\arabic*}]
 \item \emph{The first segment of \lstinbnf{FQFNElem} must be either a \rc{RCPackage} or a RoboChart module.} \label{WF:FQFNRef_head}
 \item \emph{Except for the first segment, each segment must be a child of its previous segment.} \label{WF:FQFNRef_tail}
\end{enumerate}
The example above violates both conditions \ref{WF:FQFNRef_head} and \ref{WF:FQFNRef_tail} because \lstinbnf{SRWCtrl} is not a module or a \rc{RCPackage}, and \lstinbnf{SRWMod} is not a child of \lstinbnf{SRWCtrl}.
\lstinbnf{FQFNElem} implements a mechanism to identify each instance of RoboChart elements through component references uniquely.
\lstinbnf{SRWMod::ctrl_ref::stm_ref::Move}, for example, refers to the \rc{Move} state in Fig.~\ref{fig:srw_model} through the controller reference \lstinbnf{ctrl_ref} (hidden in the figure) and the state machine reference \lstinbnf{stm_ref}. The type of a \lstinbnf{FQFNElem} is the type of its last segment, so the type of the example is the RoboChart \rc{State}.

An event is a particular named element in RoboChart and cannot be identified simply by its name because its role in a connection can be for input and output. Its direction in the connection, therefore, should be taken into account. We use \lstinbnf{pEvent} to identify an event by its name (\lstinbnf{FQFNElem}) and its role (\lstinbnf{pEventDir}, either \lstinbnf{in} or \lstinbnf{out}) in a connection.
\begin{lstlisting}[language=AssertBNF,]
pEvent    ::= FQFNElem '.' pEventDir
pEventDir ::= in | out 
pEventVal ::= pEvent '.' val  
\end{lstlisting}

For example, \lstinbnf{SRWMod::ctrl_ref::stm_ref::left.out} denotes the event \rc{left} on \rc{SRWSTM} in Fig.~\ref{fig:srw_model}, and \lstinbnf{SRWMod::SRWRP::left.in} denotes \rc{left} on the platform.  For a typed event (an input event of the form \rc{evt?x} or an output event of the form \rc{evt!e} where \rc{e} is an expression), we use \lstinbnf{pEventVal} to refer to the data carried on the event.



\subsection{Constant configurations}
\label{ssec:ppl_syntax_constconfig}


Loose constants in RoboChart models or constants declared previously can be specified using the \lstinbnf{Constants} rule, whose definition is omitted here.
%
%
We show an example to configure the constants.
\begin{example}[Constant configurations of random walk]
  \label{ex:srw_constant}
\begin{lstlisting}[language=AssertBNF,]
constants C_fair_MD10_MS20_100:
  SRWMod::SRWRP::MaxDist set to 10, 
  SRWMod::SRWRP::MaxSteps from set {20 to 100 by step 10}, and
  SRWMod::SRWRP::Pl set to 0.5
\end{lstlisting}
\end{example}

This configuration sets \rc{MaxDist} to 10, \rc{Pl} to 0.5, and \rc{MaxSteps} from a set of values starting from 20 to 100 by step 10. This is the same as nine constant configurations that set \rc{MaxSteps} to 20, 30, \ldots, and 100, respectively.

%


\subsection{Rewards}
\label{ssec:ppl_syntax_rewards}


In addition to probability, states and transitions in RoboChart models can be associated with real-valued rewards for quantitative measurements of expected values, for example, the expected number of movements after the robot returns to the origin again for the random walk. 
%
We show an example of reward definitions below.
\begin{example}[Rewards]
\begin{lstlisting}[language=AssertBNF,]
rewards R_origins = 
  [SRWMod::ctrl_ref::stm_ref::left.out] (SRWMod::SRWRP::x==0) : 1;
  [SRWMod::ctrl_ref::stm_ref::right.out] (SRWMod::SRWRP::x==0) : 1;
endrewards
\end{lstlisting}
\end{example}
The \lstinbnf{R_origins} contains two rewards, and each assigns a reward 1 to the \rc{left} or \rc{right} event when the robot is at the origin (\rc{x==0}).


\subsection{Function and operation definitions}
\label{ssec:ppl_syntax_func_op_defs}


Functions in RoboChart models are specified using preconditions and postconditions described in the rich language of Z predicates, and they can also be unspecified in the models. These predicates should not be parsed and implemented to verify the models with model checking. They provide specifications for the functions, and users must supply their implementations for verification.

Operations that are provided by robotic platforms are also unspecified in RoboChart models. The \lstinbnf{Definitions} rule below is used to specify loose functions and operations in the models.
\begin{lstlisting}[language=AssertBNF,]
Definitions ::= defs N ':' (pFunction | pOperation)+
pFunction   ::= pfunction N '(' N* ')' '=' '{' return pExpr '}'
pOperation  ::= poperation N '(' N* ')' '=' '{' pAssignment (and pAssignment)* '}'
pAssignment ::= '(' FQFNElem '=' pExpr')'
\end{lstlisting}

A name \lstinbnf{N} is associated with one or more function definitions \lstinbnf{pFunction} and operation definitions \lstinbnf{pOperation}. All loose functions and operations in a model must be defined. A function definition \lstinbnf{pFunction} has a name \lstinbnf{N}, zero or more parameters \lstinbnf{N*}, and \lstinbnf{return}s an expression \lstinbnf{pExpr}. An operation definition is similar, but it does not return an expression. Instead, it updates the variables \lstinbnf{FQFNElem} to the expression \lstinbnf{pExpr} by one or more assignments \lstinbnf{pAssignment}.  For instance, the three functions \rc{Plus}, \rc{Minus}, and \rc{Update} in the random walk model in Fig.~\ref{fig:srw_model} can be defined.
\begin{example}[Functions]
  \label{ex:srw_functions}
\begin{lstlisting}[language=AssertBNF,]
defs D_recharge: 
  pfunction Plus(v, maxv) = {...}
  pfunction Minus(v, minv) = {...}
  pfunction Update(v, maxv, origin) = { return (if $$origin then 0 else (if (($$v) < ($$maxv)) then ($$v+1) else ($$v) end) end) }
\end{lstlisting}
\end{example}
We use \lstinbnf{$$v} to refer to a parameter $v$ in a function definition to ease the parsing.



\subsection{Environment modelling}
\label{ssec:ppl_syntax_environment}


As discussed previously, RoboCertProb needs to set up a specific environmental input and check its expected behaviour for generated PRISM models.  This is implemented in RoboCertProb through probabilistic modules defined by \lstinbnf{pModules} below. Users are required to have basic knowledge of PRISM to define the correct modules.
\begin{lstlisting}[language=AssertBNF,]
pModules ::= pmodules N ':' pModule+
pModule  ::= pmodule N '{' pVariable* pCommand+ '}'
pVariable::= N ':' pType (init pExpr)? ';'
pType    ::= bool | '[' pExpr to pExpr ']'
pCommand ::= '[' pEvent? ']' pExpr '->' ((pUpdate ('&' pUpdate)*) | skip) ';'
pUpdate  ::= '(' (pExpr ':')? '@' N '=' pExpr ')'
\end{lstlisting}

A \lstinbnf{pModules} associates a name \lstinbnf{N} to one or more probabilistic modules, each \lstinbnf{pModule} of which has a name \lstinbnf{N} and contains a list of variable declarations \lstinbnf{pVariable} and one or more commands \lstinbnf{pCommand}. A variable is declared with its name \lstinbnf{N}, type \lstinbnf{pType}, optional \lstinbnf{init}ial value \lstinbnf{pExpr} specified.

\begin{figure*}[htbp!]
  \centering
  \includegraphics[scale=0.70]{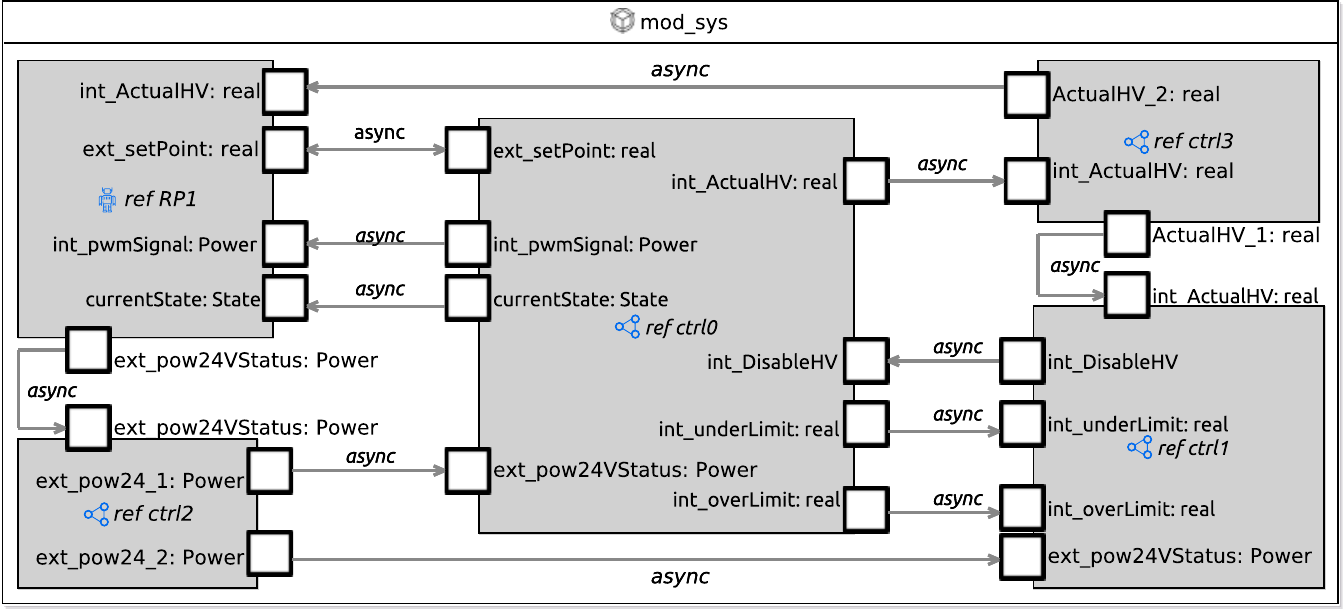}
  \caption{The HVC RoboChart module~\cite{Murray2020}: one platform \rc{RP1}, four controllers, and the connections between them.}
  \label{fig:hvc}
\end{figure*}
The use of \lstinbnf{pModules} is exemplified in specifying environmental inputs for the verification of a high voltage controller (HVC) for an industrial painting robot~\cite{Murray2020}. Figure~\ref{fig:hvc} shows its RoboChart module. 

First, we consider the requirement P3: ``That \rc{ext\_SetPoint} is set to 0 whenever the 24v power signal (\rc{ext\_pow24VStatus}) is off''. This requirement has been verified by Murray et al.~\cite{Murray2020} using FDR~\cite{T.GibsonRobinson2014}, a refinement model checker for CSP processes. P3 is specified by a CSP process \lstinrbc{Spec3} below.
\begin{lstlisting}[language=RoboCert]
Spec3 = CHAOS(Events) [| PowerOffEvents |> Follow
PowerOffEvents = {|mod_sys::ext_pow24VStatus.in.Power_Off|}
Follow = (RUN(PowerOffEvents) /\ mod_sys::ext_setPoint.out.0 -> Spec3)
\end{lstlisting}

The \lstinrbc{Spec3} initially behaves like \lstinrbc{CHAOS(Events)}, that performs any event from all events (\lstinrbc{Events}) nondeterministically until (\lstinrbc{[|...|>}, an exception operator) the occurrence of the events in \lstinrbc{PowerOffEvents}, and then behaves like the process \lstinrbc{Follow}.  \lstinrbc{PowerOffEvents} contains only one event%
, that is, the 24v power signal (\lstinrbc{ext_pow24VStatus}) to (the \lstinrbc{in} direction) the RoboChart model, is off (\lstinrbc{Power_Off}).  The \lstinrbc{Follow} process allows the further power-off events using \lstinrbc{RUN(PowerOffEvents)} until (\lstinline{/\ }, interrupt) the \emph{HV\_SetPoint} (\lstinrbc{ext_setPoint}) is observed to be 0 from (the \lstinrbc{out} direction) the model.

The CSP semantics of the RoboChart model that only exposes the events \lstinrbc{PowerOffEvents} and \lstinrbc{ext_setPoint}, then, is checked to be a trace refinement of \lstinrbc{Spec3}.

In RoboCertProb, we define a probabilistic module \lstinbnf{M3}, partially shown below. Its complete definition can be found here.\footnote{\url{https://github.com/UoY-RoboStar/hvc-case-study/blob/prism_verification/sbmf/assertions/probability_prism/P3.assertions}} In our tool, it will be generated into a PRISM module, to be in parallel composition with other modules in the PRISM model that is generated from the RoboChart model. In Markov semantics discussed in Sect.~\ref{sec:markov:robochart}, this new module introduces two new variables into its state space and more transitions. This module, eventually, sets up the environment inputs to the model for particular properties and checks the output from the model.
\begin{example}[Probabilistic modules]
  \label{example:pmodules}
\begin{lstlisting}[language=AssertBNF, literate={*}{{\(\land\)}}1,]
pmodules M3: pmodule P3 {
  // -1 - Fail, 0 - Idle, 1 - Follow
  P3_scpc: [-1 to 1] init 0;
  P3_updating: bool init false;
  // t1, t4
  [mod_sys::rp_ref0::ext_pow24VStatus.out] (@P3_updating==false)*(@P3_scpc!= -1) -> (@P3_updating=true)&(@P3_scpc=0);
  // t2
  [] (@P3_scpc==0)*(@P3_updating==true)*(mod_sys::rp_ref0::ext_pow24VStatus.out.val==Power::Off) -> (@P3_updating=false)&(@P3_scpc=1);
  // t3
  [] (@P3_scpc==0)*(@P3_updating==true)*(mod_sys::rp_ref0::ext_pow24VStatus.out.val!=Power::Off) -> (@P3_updating=false)&(@P3_scpc=0);
  ...
}
\end{lstlisting}
\end{example}

In the module \lstinbnf{P3}, we declare two variables: \lstinbnf{P3_scpc} to record the current state (from -1 to 1) of this module and a boolean \lstinbnf{P3_updating} to denote if a communication is updating its data for exchange or not. PRISM only allows synchronisation over simple actions and does not support communication that carries data directly. Additional variables, therefore, are needed for exchange. In PRISM, we need two commands (or steps) to implement a communication: first synchronisation and then exchange~\cite{Ye2022}. For this reasoning, we introduce \lstinbnf{P3_updating} to denote a communication is updating its data. For example, the command on line \verb+#6+ synchronises over the event \lstinbnf{ext_pow24VStatus.out} and then its value for exchange is accessed in the command on line \verb+#8+. We note that the direction of the event is \lstinbnf{out} now, instead of \lstinbnf{in} in \lstinrbc{Spec3}. This is because our event here is the event of the platform for output, which is equivalent to the event on the controllers for input (where \lstinrbc{Spec3} uses).

\begin{figure*}[htbp!]
  \centering
  \includegraphics[scale=0.70]{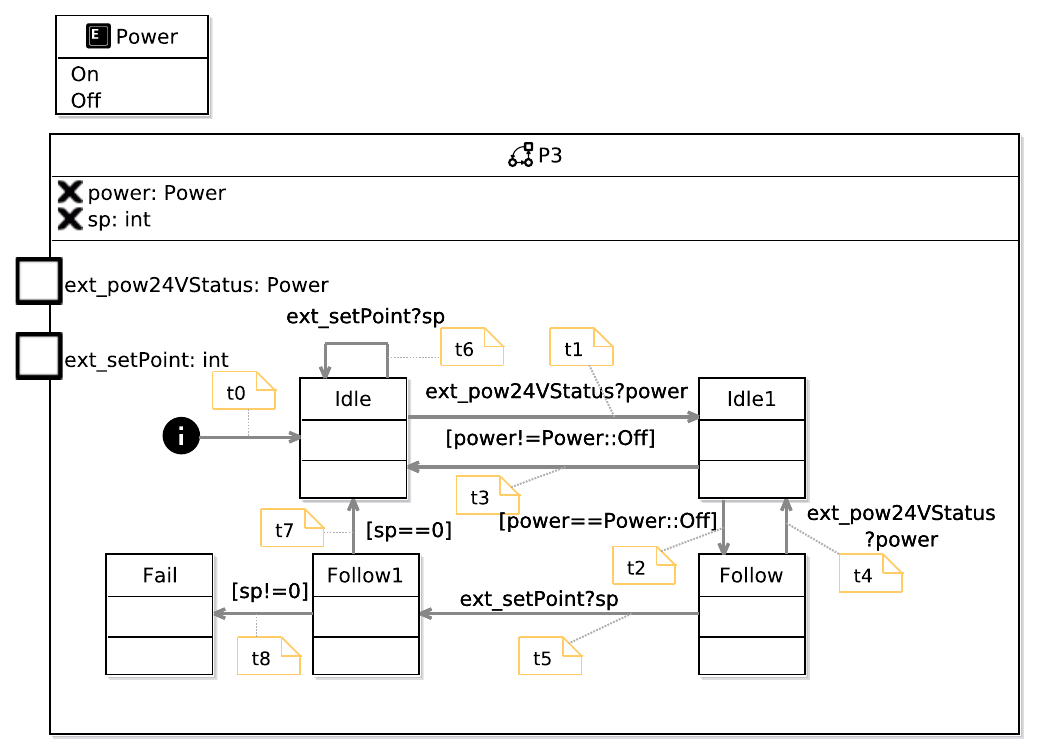}
  \caption{Illustration of a protocol implemented in the \lstinbnf{P3} module.}
  \label{fig:P3}
\end{figure*}

The \lstinbnf{P3} module above implements a protocol as illustrated in Fig.~\ref{fig:P3}, which corresponds to the process \lstinrbc{Spec3} in CSP. The mappings between the states in Fig.~\ref{fig:P3} and variables in the \lstinbnf{P3} are given in Table~\ref{tab:P3}.
\begin{table}[htbp!]
  \caption{The mappings between states and variables.}
  \label{tab:P3}
  \centering
  \begin{tabular}{@{}llllll@{}}
    \toprule
    Variables\textbackslash States & Idle & Idle1 & Follow & Follow1 & Fail \\
    \hline
    \lstinbnf{P3_scpc} & 0 & 0 & 1 & 1 & -1 \\
    \hline
    \lstinbnf{P3_updating} & False & True & False & True & False \\
    \bottomrule
  \end{tabular}
  \vspace*{-1em}
\end{table}
We use -1 for \rc{Fail}, 0 for \rc{Idle}, and 1 for the state \lstinrbc{Follow}. In Fig.~\ref{fig:P3}, \rc{Idle} corresponds to \lstinrbc{CHAOS(Events)} in \lstinrbc{Spec3}. Because \lstinbnf{P3} synchronises over only the \lstinbnf{ext_pow24VStatus} and \lstinbnf{ext_setPoint} events, it does not constrain other events from occurring. Then when a \lstinrbc{ext_pow24VStatus} event happens, the transition \rc{t1} is taken, and the state machine \rc{P3} enters the state \rc{Idle1}. If the request in the event is \rc{Power::Off} (equivalent to \lstinrbc{Power_Off}), the machine enters the state \rc{Follow} after t2 is taken. This signals that \lstinrbc{Spec3} behaves like \lstinrbc{Follow} now after the power-off event. If the request in the event is not \rc{Power::Off}, the machine returns to \rc{Idle} after t3 is taken. At the state \rc{Follow}, the machine can further read the power status and return to \rc{Idle1} after t4 is taken or read the set point and enter the state \rc{Follow1} after t5 is taken. If the set point is 0 as expected, t7 is taken, and the machine returns to \rc{Idle}. This corresponds to the interrupt of \lstinrbc{RUN(PowerOffEvents)} in the process \lstinrbc{Follow} by the \lstinbnf{ext_setPoint} event, and then \lstinrbc{Follow} behaves like \lstinrbc{Spec3}. If the set point is not 0, the machine enters the state \rc{Fail} after t8 is taken. This is unexpected behaviour.

With the \lstinbnf{P3} module, we turn a trace refinement check in FDR for \lstinrbc{Spec3} into a reachability analysis using PRISM to check if the state \rc{Fail} is not reachable.

Next, we consider P1: ``That the actual system voltage (\rc{int\_ActualHV}) always follows the set-point (\rc{ext\_setPoint})''. In CSP, its specification is split into two cases: whether the main state machine of \rc{ctrl0} is in a state \rc{ClosedLoop} or \rc{ErrorMode}, captured in two CSP specifications \lstinrbc{Spec1A} or \lstinrbc{Spec1B}. We show \lstinrbc{Spec1A} below.

\begin{lstlisting}[language=RoboCert]
Spec1A = (Behaviour [| {| int_ActualHV |} |] BufferedOutput) \ {|int_ActualHV|}
Behaviour = CHAOS(Events) [| {| 
  mod_sys::currentState.out.State_ClosedLoop |} |> 
  (Follow /\ (mod_sys::currentState.out.State_ErrorMode -> Behaviour))
Follow = mod_sys::ext_setPoint.in?x__ -> int_ActualHV!x__ -> Follow 
\end{lstlisting}
In \lstinrbc{Spec1A}, the \lstinrbc{BufferedOutput} (whose definition is omitted here and can be found online\footnote{\url{https://github.com/UoY-RoboStar/hvc-case-study/blob/prism_verification/sbmf/properties.assertions}.}) is one-place buffer which allows overriding current item in the buffer, to simulate the asynchronous connection on event \rc{int\_ActualHV} between \rc{ctrl0} and \rc{ctrl3} shown in Fig.~\ref{fig:hvc}. 

In RoboCertProb, we also use two properties \lstinbnf{P1A}\footnote{\url{https://github.com/UoY-RoboStar/hvc-case-study/blob/prism_verification/sbmf/assertions/probability_prism/P1A.assertions}.} and \lstinbnf{P1B}\footnote{\url{https://github.com/UoY-RoboStar/hvc-case-study/blob/prism_verification/sbmf/assertions/probability_prism/P1B.assertions}.} (corresponding to \lstinrbc{Spec1A} and \lstinrbc{Spec1B}) to verify P1. We show a snippet of \lstinbnf{P1A} below.

\begin{lstlisting}[language=AssertBNF, literate={*}{{\(\land\)}}1,]
pmodules M1A: 
pmodule P1A_alternate {
  send2buff: bool init true;
  [mod_sys::ctrl_ref0::stm_ref0::int_ActualHV.out] (@send2buff==true) -> (@send2buff=false);
  [mod_sys::rp_ref0::int_ActualHV.in] (@send2buff==false) -> (@send2buff=true);
}  

pmodule P1A {
	P1A_scpc : [-1 to 2] init 0;
	P1A_updating: bool init false;
...
}
\end{lstlisting}
The \lstinbnf{pmodule} \lstinbnf{P1A} is a usual module similar to the \lstinbnf{pmodule} \lstinbnf{P3} shown previously. We define another \lstinbnf{pmodule} \lstinbnf{P1A_alternate} to force an alternation between the two events: one on line \verb+#4+ for \rc{ctrl0} to write to the buffer and one on line \verb+#5+ for \rc{rp0} to read from the buffer. These events do not directly deal with the value in the buffer, but specify an environment requirement to the generated PRISM model (by the combination of the two modules with the modules in the PRISM model together). This is to simulate the asynchronous connection in the RoboChart model.

\subsection{Probabilistic properties}
\label{ssec:ppl_syntax_prob}


Qualitative and quantitative properties are specified using the \lstinbnf{ProbProperty} rule defined below.
\begin{lstlisting}[language=AssertBNF,]
ProbProperty ::= prob property N ':' pExpr
  (with constants (ConstConfig+ | N))?
  (with definitions ((pFunction|pOperation)+ | N))?
  (with modules (pModule+ | N))?/*
  (with cmdoptions STRING)?*/
\end{lstlisting}
A property associates a name \lstinbnf{N} with an expression \lstinbnf{pExpr} in various optional settings: constant configurations \lstinbnf{ConstConfig} or a reference \lstinbnf{N}, function and operation definitions (\lstinbnf{pFunction} and/or \lstinbnf{pOperation}), and probabilistic modules \lstinbnf{pModule}.

The \lstinbnf{ProbProperty} satisfies the well-formedness conditions as follows.  \smallskip
\begin{enumerate}[label=WFProp-{\arabic*}]
\item \emph{The expression must be boolean or a state formula with a query (see Sect.~\ref{ssec:ppl_syntax_pexpr}).} \label{WF:Prop_pexpr}
\item \emph{A name after \lstinbnf{with constants} must be a reference to configurations by \lstinbnf{Constants}.} \label{WF:Prop_constant_N}
\item \emph{A name after \lstinbnf{with definitions} must be a reference to function and operations by \lstinbnf{Definitions}.} \label{WF:Prop_constant_defs}
\item \emph{A name after \lstinbnf{with modules} must be a reference to modules defined by \lstinbnf{pModules}.} \label{WF:Prop_constant_pmodules}
\end{enumerate}
We illustrate it with a qualitative property to specify the random walk is deadlock-free in a specific constant configuration and a function definition.
\begin{example}[Probabilistic property]
  \label{example:probproperty}
\begin{lstlisting}[language=AssertBNF,]
prob property P_deadlock_free:
  not Exists [Finally deadlock]
  with constants C_fair_MD10_MS20_100
  with definitions D_recharge
\end{lstlisting}	
\end{example}

The expression of the property is the negation (\lstinbnf{not}) of a state formula (\lstinbnf{Exists [Finally deadlock]}) which will be described as follows.


\subsection{Expressions}
\label{ssec:ppl_syntax_pexpr}


RoboChart has a rich expression language~\cite{A.Miyazawa2020}, inspired by the Z notation. The expression language in RoboCertProb supports a subset of RoboChart's expressions, plus expressions for path and states formulas in PCTL*. Expressions are defined by the \lstinbnf{pExpr} rule below, where the grammar for literals, logical expressions, relational expressions, arithmetic expressions, conditional, and arrays is omitted here.
\begin{lstlisting}[language=AssertBNF,literate={`}{{\$}}1]
pExpr ::= /*INT | FLOAT | BOOLEAN
  | pExpr iff pExpr | pExpr '=>' pExpr
  | pExpr '\/' pExpr | pExpr '/\ ' pExpr
  | not pExpr | pExpr '==' pExpr
  | pExpr '!=' pExpr | pExpr '>' pExpr
  | pExpr '>=' pExpr | pExpr '<' pExpr
  | pExpr '<=' pExpr 
  | pExpr '+' pExpr | pExpr '-' pExpr 
  | pExpr '*' pExpr | pExpr '/' pExpr 
  | pExpr '%' pExpr | '-' pExpr
  | if pExpr then pExpr else pExpr end
  | pExpr '[' pExpr (',' pExpr)* ']'
  | */ '{' pExpr (',' pExpr)* '}'
  | '{' pExpr to pExpr (by step pExpr)?'}'
  | StateFormula | PathFormula | RPathFormula | FQFNElem
  | FQFNElem is in FQFNElem
  | '@' (N '::' N '::')? N | LabelRef | '`' N | '``' N 
  | '&' N '(' pExpr (',' pExpr)* ')' | pEventVal | '(' pExpr ')'

LabelRef ::= '#' N | deadlock | init
\end{lstlisting}

These include
\begin{enumerate*}[label={(\alph*)}]
\item set extensions on line \verb|#1|;
\item set ranges on line \verb|#2|;
\item state formulas \lstinbnf{StateFormula};
\item path formulas \lstinbnf{PathFormula};
\item reward path formulas \lstinbnf{RPathFormula};
\item references to RoboChart elements \lstinbnf{FQFNElem};
\item current state check (a composite state \lstinbnf{is in} a substate) to specify an AP for the $pc$ variable in Markov semantics, as discussed in Sect.~\ref{sec:markov:robochart};
\item references to variables in \lstinbnf{pModule} on line \verb|#5|;
\item \lstinbnf{LabelRef} (defined on line \verb|#8|): references to labels with a prefix \lstinbnf{'#'}, and pre-defined \lstinbnf{init} and \lstinbnf{deadlock};
\item references to formulas with a prefix \lstinline[language=AssertBNF,columns=fullflexible,breaklines=true,literate={`}{{\$}}1]{'`'};
\item references to parameters of function definitions \lstinbnf{pFunction} or operation definitions \lstinbnf{pOperation} with a prefix \lstinline[language=AssertBNF,columns=fullflexible,breaklines=true,literate={`}{{\$}}1]{'``'};
\item calls to defined functions \lstinbnf{pFunction} with a prefix \lstinbnf{'&'};
\item references to data on events by \lstinbnf{pEventVal}; and 
\item parenthesised expressions.
\end{enumerate*}

We show an example below for a label \lstinbnf{l_stuck} for the specification if the current state of the machine \rc{SRWSTM} in Fig.~\ref{fig:srw_model} is \rc{Stuck} and a label \lstinbnf{l_origin} for the specification if the value of \rc{x} is equal to 0. 
\begin{example}[Expressions]
  \label{example:pexpr}
\begin{lstlisting}[language=AssertBNF,]
label l_stuck = 
	SRWMod::ctrl_ref::stm_ref is in SRWMod::ctrl_ref::stm_ref::Stuck
label l_origin = (SRWMod::SRWRP::x == 0)
\end{lstlisting}
\end{example}
In Markov semantics, they actually specify two APs \lstinbnf{(pc=Stuck)} and \lstinbnf{(x=0)}, as shown in Sect.~\ref{sec:markov:robochart}.

State formulas, as given in the syntax of PCTL* in Sect.~\ref{sec:markov:pctl}, specify properties of states, defined by \lstinbnf{StateFormula} below.
\begin{lstlisting}[language=AssertBNF,morekeywords={Reward}]
StateFormula ::= PFormula | RFormula | AFormula | EFormula
PFormula     ::= Prob (Bound | Query) of '[' pExpr ']' (UseMethod)?
RFormula     ::= Reward ('{' N '}')? (Bound | Query) of '[' RPathFormula ']' (UseMethod)?
AFormula     ::= Forall '[' pExpr ']'
EFormula     ::= Exists '[' pExpr ']'
Bound ::= ('>' | '>=' | '<' | '<=') pExpr
Query ::= '?=' | min '?=' | max '?=' 
\end{lstlisting}

State formulas include
\begin{enumerate*}[label={(\alph*)}]
\item probability formulas \lstinbnf{PFormula}: either \lstinbnf{Bound} (probability compared with an expression using one of four comparison operators) or \lstinbnf{Query} (what is the probability, the minimum probability, or the maximum probability), optionally using simulation methods \lstinbnf{UseMethod} whose definition is omitted for simplicity;
\item reward formulas \lstinbnf{RFormula} with optional references \lstinbnf{N} to defined \lstinbnf{rewards}: either \lstinbnf{Bound} or \lstinbnf{Query}; and
\item non-probabilistic path quantifiers: for all paths \lstinbnf{AFormula} and for some paths \lstinbnf{EFormula}.
\end{enumerate*}

State formulas are the probabilistic operator (P) and non-probabilistic operators (A and E) over expressions.  Path formulas \lstinbnf{PathFormula} are defined as follows.

\begin{lstlisting}[language=AssertBNF,]
PathFormula ::= Next pExpr
  | pExpr Until (Bound)? pExpr
  | Finally (Bound)? pExpr
  | Globally (Bound)? pExpr
  | Weak Until (Bound)? pExpr
  | Release (Bound)? pExpr
\end{lstlisting}

Path formulas include 
\begin{enumerate*}[label={(\alph*)}]
\item \lstinbnf{Next} (X), 
\item \lstinbnf{Until} (U), 
\item \lstinbnf{Finally} (F), 
\item \lstinbnf{Globally} (G), 
\item \lstinbnf{Weak Until} (W), and
\item \lstinbnf{Release} (R).
\end{enumerate*}
Except \lstinbnf{Next}, other path formulas can also be bounded in terms of a given number of steps, such as \lstinbnf{Finally<=10}. We note that the time steps here are expressed in the underlying discrete Markov Chain and not in RoboChart's time semantics. For this reason, \lstinbnf{Next} and bounded time steps have no RoboChart interpretation. We, however, could explicitly model discrete time in RoboChart models using, for example, an additional variable of type natural numbers. And then, we can specify properties over this variable to quantify time. It is also worth mentioning that \lstinbnf{Next} and bounded path formulas are still useful, to some extent, for example, in verifying reachability in inaccurate bounded time steps, though they do not have a RoboChart interpretation.

The reward operator (R) is over particular path formulas \lstinbnf{RPathFormula}.
\begin{lstlisting}[language=AssertBNF,]
RPathFormula ::= Reachable pExpr | LTL pExpr | Cumul pExpr | Total
\end{lstlisting}
Reward path formulas include reachability rewards (\lstinbnf{Reachable}), co-safe \lstinbnf{LTL} rewards that only F, U, and X operators are used in the subsequent \lstinbnf{pExpr} and not the G operator, \lstinbnf{Cumul}ative rewards, and \lstinbnf{Total} rewards.\footnote{\url{www.prismmodelchecker.org/manual/PropertySpecification/Reward-basedProperties}}

Expressions are categorised into boolean or non-boolean expressions, numerical or non-numerical expressions, and set or non-set expressions.

Boolean expressions include literal \lstinbnf{true} and \lstinbnf{false}, logical, relational, conditional expressions if both expressions of the \lstinbnf{then} and \lstinbnf{else} branches are boolean, array expressions if their basic types are boolean, state formulas if they do not contain queries, path formulas, reward path formulas, references to RoboChart variables or \lstinbnf{pModule} variables that are boolean, current state check, references to labels, references to the values of events that are of type boolean, references to formulas that are boolean, and calls to \lstinbnf{pFunction}s whose results are boolean. The other expressions are non-boolean.

All boolean expressions are non-numerical expressions. Numerical expressions include integers, real numbers, arithmetic expressions, conditional expressions if both expressions of the \lstinbnf{then} and \lstinbnf{else} branches are numerical, array expressions if their basic types are numerical, state formulas if they are queries, references to RoboChart variables or \lstinbnf{pModule} variables that are numerical, references to the values of events that are numerical, references to formulas that are numerical, and calls to \lstinbnf{pFunction}s whose results are numerical. The other expressions are non-numerical.

Set extension and range expressions are set expressions, and others are non-set expressions.

The \lstinbnf{pExpr} satisfies the well-formedness conditions partially shown below.
\begin{enumerate}[label=WFExp-{\arabic*}]
\item \emph{Expressions in logical expressions must be boolean.} \label{WF:Expr_logical}
\item \emph{Expressions of equality and inequality must be the same type.} \label{WF:Expr_equation}
\item \emph{Expressions in comparison and arithmetic expressions must be numerical.} \label{WF:Expr_relational_arithmetic}
 \item \emph{Expressions in set extensions must have the same type.} \label{WF:Expr_set_ext}
\item \emph{The first reference in the current state check must be a state machine or a composite state, and the second must be its immediate substate.} \label{WF:Expr_is_in}
\item \emph{The expression in \lstinbnf{PFormula}, \lstinbnf{AFormula}, or \lstinbnf{EFormula} must be \lstinbnf{PathFormula}.} \label{WF:Expr_PathFormula}
\item \emph{The expression in \lstinbnf{Bound} must be numeric.} \label{WF:Expr_Bound}
\end{enumerate}

 \subsection{Simulation}
 \label{ssec:ppl_syntax_sim}
 \vspace*{-0.5em} In addition to probabilistic model checking (verification), both probability and reward operators also support statistical model checking\footnote{\url{www.prismmodelchecker.org/manual/RunningPRISM/StatisticalModelChecking}} which uses sample-based discrete-event simulation. 
 \lstinbnf{UseMethod} is provided to configure simulation methods.
\begin{lstlisting}[language=AssertBNF,]
UseMethod ::= using sim with SimMethod (',' and pathlen '=' pExpr)?
SimMethod ::= CI (at CiMethod)?  | ACI (at CiMethod)?  | APMC (at APMCMethod)?  | SPRT (at SPRTMethod)?
CiMethod ::= ((',')? w '=' pExpr)?  & ((',')? alpha '=' pExpr)?  & ((',')? n '=' pExpr)?
APMCMethod ::= ((',')? epsilon '=' pExpr)?  & ((',')? delta '=' pExpr)? & ((',')? n '=' pExpr)?
SPRTMethod ::= ((',')? alpha '=' pExpr)?  & ((',')? delta '=' pExpr)?
\end{lstlisting}
 
 A simulation method is specified in \lstinbnf{SimMethod} with an optional \lstinbnf{pathlen} to specify the maximum length of simulation paths. There are four configurable simulation methods: Confidence Interval (\lstinbnf{CI}), Asymptotic Confidence Interval (\lstinbnf{ACI}), Approximate Probabilistic Model Checking (\lstinbnf{APMC}), and Sequential Probability Ratio Test (\lstinbnf{SPRT}). The symbol \lstinbnf{&} is separate for unordered groups in Xtext. That is, these groups can come in any order.

%
 We illustrate the use of simulation to verify P5 of HVC~\cite{Murray2020}: that all states in the state machine are reachable.
\begin{example}[Simulation]
  \label{example:simulation}
\begin{lstlisting}[language=AssertBNF,]
prob property P5_reach_ClosedLoop:
   Prob>0 of [Finally mod_sys::ctrl_ref0::stm_ref0 is in mod_sys::ctrl_ref0::stm_ref0::ClosedLoop]
   using sim with CI at alpha=0.01, n=100, and pathlen=1000000
   with definitions D1
\end{lstlisting}	
\end{example}
 We use the confidence interval method with confidence (\lstinbnf{alpha}) 0.01, the number of samples (100), and the maximum length of paths (1000000) to verify there is a non-zero probability (\lstinbnf{>0} that the state machine \lstinbnf{stm_ref0} is \lstinbnf{Finally} in its substate \lstinbnf{ClosedLoop}).

\section{Tool support}
\label{sec:tool}

\begin{figure*}[tb!]
  \centering
  \includegraphics[scale=0.50]{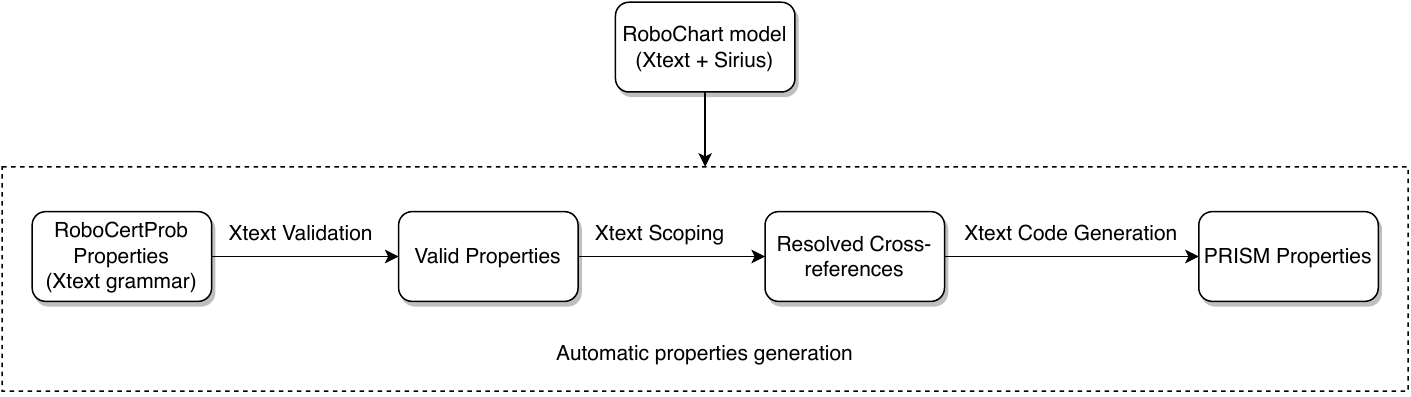}
  \caption{The procedure to generate PRISM properties from RoboCertProb using Xtext.}
  \vspace*{-1em}
  \label{fig:tool_support}
\end{figure*}

RoboTool\footnote{\url{www.cs.york.ac.uk/robostar/robotool/}} contains a collection of Eclipse plug-ins which are based on the Eclipse Modeling Framework (EMF)\footnote{\url{www.eclipse.org/modeling/emf/}} and particularly implemented using Xtext\footnote{\url{www.eclipse.org/xtext}.} and Sirius.\footnote{\url{www.eclipse.org/sirius}.} It supports modelling, validation, and automatic generation of mathematical definitions of RoboChart models written in CSP and PRISM for verification using FDR and PRISM. Our work described here extends RoboTool to model and validate RoboCertProb, and automatically generate PRISM properties from RoboCertProb for probabilistic model checking with PRISM. We describe the generation procedure in Fig.~\ref{fig:tool_support}.

\paragraph{Syntax}
The syntax of RoboCertProb is defined using the grammar language of Xtext with references to RoboChart's textual syntax (Xtext) and metamodel. We illustrate the implementation of \lstinbnf{FQFNElem} defined in Sect.~\ref{ssec:ppl_syntax_fqfnelem} below.
%
\begin{lstlisting}[language=Xtext]
FQFNElem returns QNRef:
  NameRef ({FQFNElem.ref=current} '::' tail=[NamedElement])* ;
NameRef returns QNRef:
  {RCRef} ':' rc=[RCPackage] |
  {NameRef} name=[NamedElement] ;
\end{lstlisting}
%
%
\lstinline{FQFNElem} is composed of a head (by the rule \lstinline{NameRef}) and zero or more (\lstinline{*}) tails of type \lstinbnf{NamedElement}. Both \lstinline{FQFNElem} and \lstinline{NameRef} return the same abstract type \lstinline{QNRef}, but they have different concrete types specified by actions (inside curly brackets) in Xtext: type \lstinxtx{FQFNElem}, \lstinxtx{RCRef}, and \lstinxtx{NameRef} for the rule \lstinxtx{FQFNElem} and \lstinxtx{NameRef} respectively if the head is a \lstinxtx{RCPackage} or \lstinxtx{NamedElement}.
The rule \lstinline{FQFNElem}, therefore, will instantiate an object of type \lstinline{FQFNElem} because of the tree rewrite action \lstinxtx{FQFNElem.ref=current}. The rule \lstinline{NameRef} will instantiate an object of type \lstinline{RCRef} (because of the action \lstinline|{RCRef}|) if the head is of type \lstinbnf{RCPackage} (referred by \lstinline{rc}) on line \verb+#4+ and an object of type \lstinline{NameRef} (because of the action \lstinline|{NameRef}|) if the head is of type \lstinbnf{NamedElement} (referred by \lstinline{name}) on line \verb+#5+. 
On line \verb+#2+ a special \lstinxtx{current} variable represents the element currently to-be-returned, the front part of \lstinline{FQFNElem} in this case. Considering \lstinbnf{SRWMod::ctrl_ref::stm_ref} to the state machine in Fig.~\ref{fig:srw_model}, \lstinxtx{current} denotes \lstinbnf{SRWMod} when parsing \lstinbnf{SRWMod::ctrl_ref} (\lstinline{tail} is \lstinbnf{ctrl_ref}) and denotes \lstinbnf{SRWMod::ctrl_ref} when parsing the whole \lstinbnf{SRWMod:: ctrl_ref::stm_ref} (\lstinline{tail} is \lstinbnf{stm_ref}).

\paragraph{Validation}
The RoboCertProb grammar, presented in Sect.~\ref{sec:ppl_syntax} as production rules, is flexible, but not all syntax satisfying these rules is valid. The invalid syntax is ruled out by well-formedness conditions defined for RoboCertProb. We show the conditions for \lstinbnf{FQFNElem} and \lstinbnf{pExpr} previously and omit others for simplicity.

The well-formedness conditions are implemented in RoboTool using the validation of Xtext. 
%
We show below an example of the implementation for \ref{WF:Prop_pexpr}.
%
\begin{lstlisting}[language = Xtend]
@Check def checkProbProperty(ProbProperty pa) {
  if(pa.form !== null && !isBooleanExpr(pa.form) 
     && !isQueryExpr(pa.form)) 
  { error(...) }
}
\end{lstlisting}
%
{The code uses Xtend\footnote{\url{www.eclipse.org/xtend}}, a dialect of Java. The method \lstinxtd{checkProbProperty} has a parameter \lstinxtd{pa} of type \lstinline{ProbProperty}. The \lstinxtd{pa} has a member variable \lstinxtd{form} of type \lstinbnf{pExpr}, the expression to be verified. We ensure an error is triggered if \lstinxtd{form} is not (\lstinxtd{!}) boolean (\lstinxtd{isBooleanExpr}) and not a state formula with a query (\lstinxtd{isQueryExpr}) on lines \verb+#2-3+ 
where the concrete error message is omitted for simplicity.
\paragraph{Scoping}
After validation, we get valid properties, illustrated in Fig.~\ref{fig:tool_support}. The cross-references in the properties are then resolved using the scoping of Xtext, so we know which elements are referable by a cross-reference. For example, 
the controller \lstinbnf{SRWCtrl} can now be referred to using its cross-reference \lstinbnf{SRWMod::ctrl_ref}. 

The code below illustrates part of the scoping for \lstinbnf{FQFNElem}.
%
\begin{lstlisting}[language = Xtend]
override getScope(EObject context, EReference reference) {
  ...
  else if (context instanceof FQFNElem) {
    val front = context.ref; 
    switch (front) { 
      NameRef : {
        val name = front.name
        if(name instanceof RCModule) {...} ...
      }
      RCRef : { 
        return scopeFor(front.rc.modules, NULLSCOPE) 
      }
      FQFNElem : { 
        val tail = front.tail
        switch (tail) {
          ControllerRef: {
            return tail.ref.elementsDeclared(NULLSCOPE)
          }
          ...  
      }}}} 
}
\end{lstlisting}


%
The \lstinxtd{context} is of type \lstinxtd{FQFNElem} (on line \verb+#3+). The currently processing reference like \lstinbnf{SRWMod::ctrl_ref}, is denoted by \lstinxtd{context.ref} or \lstinxtd{front} (on line \verb+#4+). Depending on the type of \lstinxtd{front}, different scopings are implemented: lines \verb+#6-9+ for \lstinxtd{NamedRef}, lines \verb+#10-12+ for \lstinxtd{RCRef}, and lines \verb+#13-20+ for \lstinxtd{FQFNElem}. The implementation procedure is similar:
\begin{enumerate*}[label={(\alph*)}]
\item get the last reference in \lstinxtd{front}: \lstinxtd{name} for \lstinxtd{NamedRef}, \lstinxtd{rc} for \lstinxtd{RCRef}, and \lstinxtd{tail} for \lstinxtd{FQFNElem}, such as \lstinbnf{ctrl_ref}; and
\item recursively call scoping for children of the identified last reference, depending on its type (RoboChart modules on line \verb+#8+ and controller references on lines \verb+#16-18+ and the subsequent call of \lstinxtd{elementsDeclared} for the scoping of children of referred \lstinxtd{ControllerDef}).
\end{enumerate*}
For example, the reference \lstinbnf{SRWMod::ctrl_ref} is valid because \lstinbnf{ctrl_ref} is a node of the module \lstinbnf{SRWMod} and so whose scoping is called on line \verb+#8+.

\paragraph{Code generation}
Finally, we use Xtext code generation to produce PRISM properties from RoboCertProb. 
For example, the snippet below shows the code generation of the \lstinbnf{Finally} operator.
%
\begin{lstlisting}[language = Xtend,extendedchars=true, literate={«}{{\(\ll\)}}1 {»}{{\(\gg\)}}1,]
def dispatch CharSequence compilePathFormula(FinalFormula e) {
  return '''(F <<IF e.b !== null>><<e.b.compile>><<ENDIF>>
            <<e.f.compileExpr>>)'''
}
\end{lstlisting}
%
 The counterpart of \lstinbnf{Finally} in PRISM is \lstinxtd{F}, followed by the compilation of the bound \lstinxtd{e.b} if there is one and the expression \lstinxtd{e.f}.
We also note that this complete generation procedure needs to take RoboChart models as input because properties are specified, validated, and scoped in terms of RoboChart models.

\section{Verification of robotic examples}
\label{sec:examples}

\subsection{Simple random walk}
For the simple random walk in Fig.~\ref{fig:srw_model}, we consider the constant configurations in Example~\ref{ex:srw_constant} for a fair coin, the function definitions in Example~\ref{ex:srw_functions} for recharging at the origin where \rc{steps} is reset to 0 by \rc{Update}. We explored biased coins where \lstinbnf{SRWMod::SRWRP::Pl from set \{0.3, 0.8\}} and \rc{Update} without returning 0, and so \rc{steps} is not reset (without recharging).

The interesting properties that we have verified include deadlock-freedom shown in Example~\ref{example:probproperty}, the probability that the random walk gets \rc{Stuck} at places other than its origin, and the expected number of times that it returns to its origin before getting \rc{Stuck}. The last two properties or queries are shown below.
\begin{lstlisting}[language=AssertBNF,]
label l_stuck = SRWMod::ctrl_ref::stm_ref is in SRWMod::ctrl_ref::stm_ref::Stuck
label l_origin = (SRWMod::SRWRP::x == 0)
prob property P_stuck_not_origin:
  Prob=? of [Finally #l_stuck /\ not #l_origin ]
  ...
prob property R_stuck_not_origin:
  Reward {R_origins} =? of [Reachable #l_stuck /\ not #l_origin ]
  ...
\end{lstlisting}

Two labels are defined on lines \verb+#1-2+ and used on lines \verb+#4+ and \verb+#7+. 

The RoboChart model, the properties verified, and the verification result for this example can be found on the RoboStar website.\footnote{\url{robostar.cs.york.ac.uk/prob_case_studies/Simple_Random_walk/index.html}} The verification result shows the model is deadlock-free under the constant configurations for fair and biased coins and the function definitions for both with and without recharging. And the probability of getting stuck is 1 for all configurations and definitions. 

The results of \lstinbnf{R_stuck_not_origin} are listed in Table~\ref{tab:srw_rewards}. We can see that a larger \lstinbnf{MaxSteps} means the more times it returns to its origin and that the expected times with recharging on row 3 are longer than those without recharging on row 2. We also note that a bigger probability difference (0.4 for Pl=0.3 and 0.6 for Pl=0.8) between left and right leads to a smaller expected number of times. The random walk is less likely to return to its origin using a more biased coin. For Pl=0.8, the results are stationary on row 5, which means the random walk is more likely to get stuck before 20 steps.

\begin{table}[tbp]
  \caption{The expected number of return times.}
  \label{tab:srw_rewards}
  \centering
\bgroup
\def\arraystretch{1.2}
\setlength\tabcolsep{.8mm}
  \begin{tabular}{@{}lcccccc@{}}
    \toprule
    \multirow{2}{*}{Configuration} & & \multicolumn{5}{c}{MaxSteps} \\
    \cmidrule{3-7}
     & & 20 & 40 & 60 & 80 & 100 \\
    \midrule
    Pl=0.5 and non-recharging & & 2.5 & 4.0 & 5.1 & 6.2 & 7.3 \\
    \midrule
    Pl=0.5 and recharging & & 4.7 & 7.0 & 9.5 & 12.5 & 16.2 \\
    \midrule
    Pl=0.3 and non-recharging && 1.33 & 1.47 & 1.5 & 1.51 & 1.7 \\
    \midrule
    Pl=0.8 and non-recharging && 0.66 & 0.67 & 0.67 & 0.67 & 0.67 \\
    \bottomrule
  \end{tabular}
\egroup
  \vspace*{-1em}
\end{table}

\subsection{High Voltage Controller (HVC)}
For the HVC example~\cite{Murray2020} used in an industrial paint robot, its RoboChart model was previously verified for safety assurance based on the CSP semantics of RoboChart using FDR. We have also verified all five properties P1 to P5, based on their probabilistic semantics using PRISM. This model is not probabilistic, and all properties are qualitative. These properties involve the expected outputs from the model based on particular inputs, such as P1 ``that the actual system voltage always follows the set point''. As discussed, this is not directly supported in PRISM verification due to the closed-world assumption. We use RoboCertProb to define extra PRISM modules\footnote{\url{https://github.com/UoY-RoboStar/hvc-case-study/blob/prism_verification/sbmf/assertions/probability_prism/}} to specify the environmental inputs and turn trace refinement check in FDR into reachability check in PRISM.

In Example~\ref{example:pmodules}, we show the module \lstinbnf{M3} for \lstinbnf{P3}, and the probabilistic properties for verification below. 
%
\begin{lstlisting}[language=AssertBNF,]
label l3 = @M3::P3::P3_scpc != -1

prob property P3_deadlock_free:
  not Exists [Finally deadlock]
  with definitions D1
  with modules M3

prob property P3:
  Forall [Globally #l3]
  with definitions D1
  with modules M3
\end{lstlisting}
The label \lstinbnf{l3} corresponds to a condition that the \rc{Fail} state in Table~\ref{tab:P3} or Fig.~\ref{fig:P3} is not reached. The property \lstinbnf{P3_deadlock_free} on line \verb+#3+ specifies that the composition of the generated PRISM model from its RoboChart model with this module \lstinbnf{M3} is deadlock-free. The property \lstinbnf{P3} on line \verb+#8+ specifies that the \rc{Fail} state is never reachable. Both are verified to be true. 

Our verification using PRISM shows all five properties are satisfied, which is the same as that of~\cite{Murray2020} using FDR (based on its CSP semantics). In Table~\ref{tab:hvc_performance}, we compare the performance of our verification using PRISM with FDR. The results show that the state space (the number of states and transitions) of the generated PRISM model is much larger than the generated and optimised CSP model with the compressions: strong bisimulation and diamond elimination~\cite{Roscoe2011}. This is due to several factors: 
\begin{enumerate*}[label=(\arabic*)]
    \item the extra intermediate states (see Figs.~\ref{fig:robochart_trans_to_markov_trans} and \ref{fig:robochart_trans_to_markov_trans_comms}) we need to introduce in our Markov semantics; 
    \item the PRISM semantics~\cite{PRISMTeam2008} is not compositional, and 
    \item PRISM does not support a similar compression technique to reduce its state space.
\end{enumerate*}
This raises a research question about how to effectively use abstraction (including variable elimination) to reduce the state space of Markov models. In our work, the discrete time in Markov models is not used (as explained in Sect.~\ref{sec:markov:motivation}) and many intermediate states can be eliminated because they will not be quantified in RoboCertProb. However, it is still a challenge to eliminate these states without an impact on the distributions of other interesting Markov states (e.g. those corresponding to RoboChart states). This is part of our future work.

The verification of the properties P1 to P4 using PRISM starts with the model construction (CnT, to compile PRISM models into Markov models) and then model checking (ChT). The overall time spent for each property using PRISM is much longer than that using FDR because PRISM has an extended model construction time. However, the PRISM model checking time is just a few seconds, less than the overall model checking time (about 10 seconds) of FDR for each property. We also note that P5 specifies the reachability of RoboChart states, so we use SMC or simulation in PRISM to verify it. SMC uses sampling for approximation and is not subject to the usual state space explosion problem. It, therefore, can be used to verify large models using approximation. And these models might not have been tackled by FDR. FDR employs better state space compression techniques for this non-probabilistic model to reduce the verification complexity. FDR, however, is not able to analyse probabilistic RoboChart models, as illustrated in the following example.

\begin{table*}[tbp]
  \caption{HVC verification result using PRISM and FDR.}
  \label{tab:hvc_performance}
  \centering
\bgroup
\def\arraystretch{1.2}
\setlength\tabcolsep{.8mm}
  \begin{tabular}{@{}cc cccc c ccccc @{}}
    \toprule
    \multirow{2}{*}{Property} &\phantom{a} &  \multicolumn{4}{c}{PRISM} & \phantom{a} & \multicolumn{4}{c}{FDR (CSP)} 
    \\ 
    \cmidrule{3-6} \cmidrule{8-11}
    & & NoS& NoT & CnT[s] & ChT[s] & & NoS& NoT & CmT[s] & ChT[s] \\
    \midrule
    P1A & & 6.9e8 & 3.0e9 & \SInum{9794.881} & 1.219 & & \SInum{2152} & \SInum{15775} & 1.35 & 15.70 \\
    \midrule
    P1B && 7.6e8 & 3.3e9 & 16971.342 & 1.357& &\SInum{23107} & \SInum{295249} & 1.75 & 13.11 \\
    \midrule
    P2 && 8.0e9 & 3.8e10 & 13321.339 & 6.188 & & \SInum{16098} & \SInum{166277}&1.58 & 10.92 \\
    \midrule
    P3 && 5.1e9 & 2.4e10 & 5944.112 & 2.517 & & \SInum{20186}&\SInum{213370} &1.51 & 15.28\\
    \midrule
    P4 && 2.2e9 & 9.3e9 & 3631.007 & 1.973 & & \SInum{22767}&\SInum{336103}& 23.21&2.45 \\
    \midrule
    P5 && NA & NA & NA & 3.209 & & \SInum{67}&\SInum{257} & 0.03 & 15.66 \\
    \midrule
\multicolumn{12}{p{.62\linewidth}}{\textbf{Acronym}: 
    NoS: number of states $\ast$
    NoT: number of transitions$\ast$
    CnT: construction time$\ast$ 
    ChT: model-checking time$\ast$
    CmT: compile time $\ast$ 
        } \\
    \bottomrule
  \end{tabular}
\egroup
\end{table*}

\subsection{UV-light treatment robot (UVC)}
In agriculture, we also conducted another hazard analysis~\cite{Adam2023} for a UV-light treatment robot~\cite{Guevara2021} for plants to analyse risks during row transition. 
We started with writing a model in PRISM directly and found it was a challenge to design a correct PRISM model to capture asynchronous behaviour between different parties. And even a correct PRISM model is created, but its connection with our high-level design using state machine diagrams is loose, so a correct implementation highly relies on persons. This is an error-prone process if the design is updated later. Then we turned to using RoboChart to capture our design. This process was very straightforward, and even knowledge of PRISM is not desired, thanks to the automatic PRISM model and RoboCertProb property generation in RoboTool. 

We capture the behaviour and uncertainty of humans, an object detection system (ODS), and the robot and the interactions between them in a RoboChart model, illustrated in Figs.~\ref{fig:uvc_model:system}, \ref{fig:uvc_model:event}, and \ref{fig:uvc_model:ods}.\footnote{The complete model can be found at \url{https://github.com/UoY-RoboStar/uvc-case-study/tree/main}} 
\begin{figure*}[tb]
  \centering
  \includegraphics[scale=0.50]{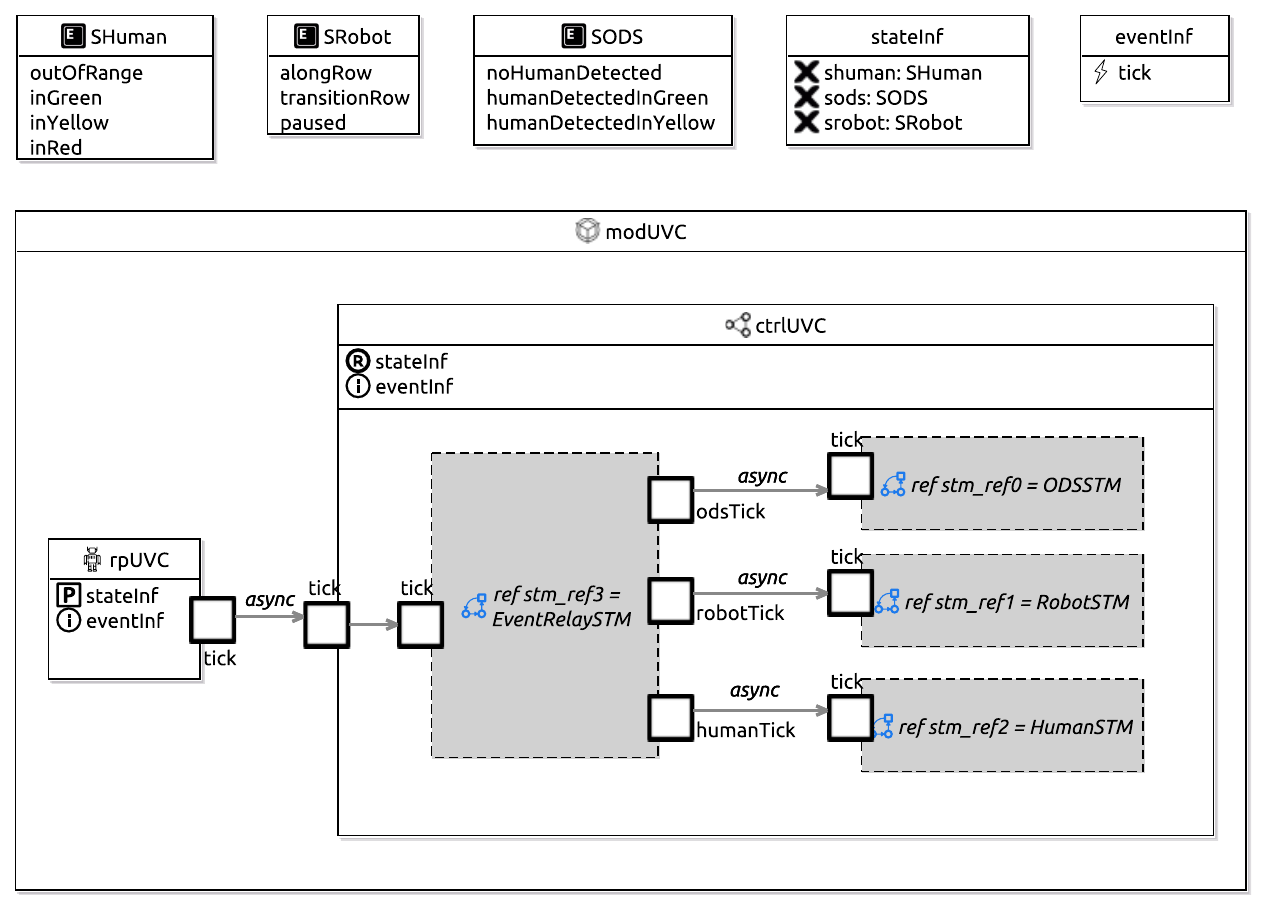}
  \caption{The UVC RoboChart model: the declarations of three enumerations (\rc{SHuman}, \rc{sRobot}, and \rc{SDOS}), one data interface \rc{stateInf}, and one event interface \rc{eventInf}; the module \rc{modUVC}, the plaform \rc{rpUVC}, and the controller \rc{ctrlUVC} with the connections between its four state machine references, from~\cite{Adam2023}.}
  \vspace*{-1em}
  \label{fig:uvc_model:system}
\end{figure*}
\begin{figure}[tb]
  \centering
  \includegraphics[scale=0.50]{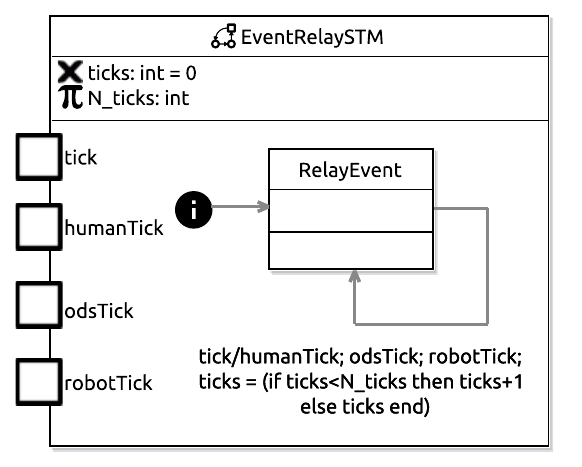}
  \caption{UVC RoboChart model: event relay state machine to keep generating tick messages until \rc{N\_ticks} is reached, from~\cite{Adam2023}.}
  \vspace*{-1em}
  \label{fig:uvc_model:event}
\end{figure}
\begin{figure*}[tb]
  \centering
  \includegraphics[scale=0.60]{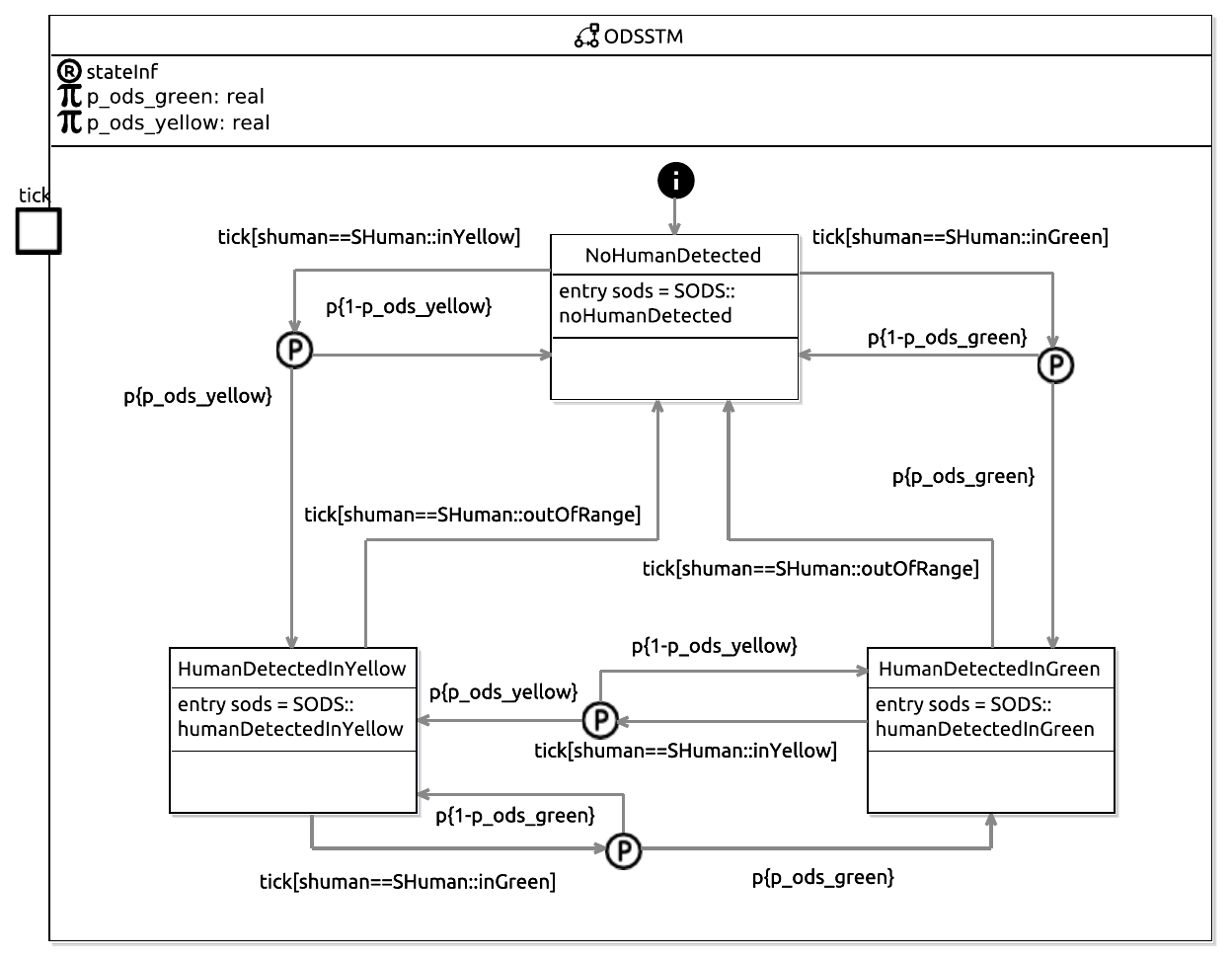}
  \caption{UVC RoboChart model: ODS state machine, from~\cite{Adam2023}.}
  \vspace*{-1em}
  \label{fig:uvc_model:ods}
\end{figure*}
Particularly, we model a global \rc{tick} event and split them into three events (\rc{humanTick}, \rc{odsTick}, and \rc{robotTick}) to allow the humans, the ODS, and the robot asynchronously align with the global tick. We use a state machine \rc{EventRelaySTM}, defined in Fig.~\ref{fig:uvc_model:event}, for the event split and declare a variable \rc{ticks} to record the number of occurred ticks, which allows us to specify properties over discrete time. This is to model these components' asynchronous nature and interleaving behaviour. Figure~\ref{fig:uvc_model:ods} captures the behaviour of the ODS in three states and the transitions between them are determinated by human position and system detection accuracy \rc{p\_ods\_green} and \rc{p\_ods\_yellow}.

We consider twelve scenarios where humans have a different safety awareness when approaching the robot, and the ODS has different accuracies. These are probability information and loose constants of the model. We use RoboCertProb to configure and specify properties for 12 different scenarios.\footnote{\url{https://github.com/UoY-RoboStar/uvc-case-study/blob/bf93d6df9380f5a644f29293898585e85ced5f1c/Properties/config.assertions}} We show the definition of one scenario \lstinbnf{C20_badOds_delibret} below where the ODS has low accuracy, and humans are delibrately approaching the robot.
\begin{lstlisting}[language=AssertBNF,]
const t : core::int
constants C20_badOds_delibret:
	modUVC::ctrlUVC::stm_ref0::p_ods_green set to 0.4,
	modUVC::ctrlUVC::stm_ref0::p_ods_yellow set to 0.7,
	modUVC::ctrlUVC::stm_ref1::p_transition_ratio set to 0.1,
	modUVC::ctrlUVC::stm_ref2::p_approach_robot set to 1,
	modUVC::ctrlUVC::stm_ref2::p_approach_yellow set to 1,
	modUVC::ctrlUVC::stm_ref2::p_approach_red set to 1, 
	modUVC::ctrlUVC::stm_ref2::p_aware_of_risk set to 0.01, 
	modUVC::ctrlUVC::stm_ref3::N_ticks set to 20, and
	t from set {1 to 19 by step 1}
\end{lstlisting}
Because the value of \lstinbnf{t} is chosen from a set of 19 elements, this scenario actually defines 19 configurations.
We are interested in the probabilities of raising a hazard where humans are in the red zone when the robot is doing a row transition in terms of discrete time. This is specified as a property below.
\begin{lstlisting}[language=AssertBNF,]
prob property P_dangerous_C20_badOds_delibret:
	Prob=? of [Finally modUVC::rpUVC::shuman==SHuman::inRed /\ 
		modUVC::rpUVC::srobot==SRobot::transitionRow /\
		modUVC::ctrlUVC::stm_ref3::ticks==t
	]
	with constants C20_badOds_delibret
\end{lstlisting}
Altogether, we configure 12 scenarios in 150 lines of property specification and specify one property for each scenario in 90 lines. We analysed 229 properties using RoboTool and the verification took just a few minutes.

\subsection{Discussions}
In Table~\ref{tab:matrix}, we show the number of properties verified for the three examples, and the number of lines in RoboCertProb to configure loose constants and function definitions, and 
 to specify these poperties. For the three examples, it takes less than 3 seconds for RoboTool to generate the corresponding PRISM properties from RoboCertProb properties.

\begin{table*}[tbp]
  \caption{General measurements of three examples.}
  \label{tab:matrix}
  \centering
\bgroup
\def\arraystretch{1.2}
\setlength\tabcolsep{.8mm}
  \begin{tabular}{@{}cc cc cc c @{}}
    \toprule
    Item &\phantom{a} & SRW  &\phantom{a} & HVC &\phantom{a} & UVC \\
    \midrule
    Number of properties & & 127 & & 9 & & 229 \\
    \midrule
    Number of lines in RoboCertProb (configurations) & & 39 & & 10  & & 150 \\ 
    \midrule
    Number of lines in RoboCertProb (properties) & & 61 & & 340 & & 90 \\
    \midrule
    Time to generate PRISM properties & \multicolumn{6}{c}{< 3 seconds} \\
    \bottomrule
  \end{tabular}
\egroup
\end{table*}

\section{Related work}
\label{sec:related}

A property specification patterns (PSP) system is presented by Dwyer et al.~\cite{Dwyer1999} to facilitate users to write property specifications (usually in LTL and CTL) for model-checkers and finite-state verification tools. 
Based on PSP, Smith et al.~\cite{Smith2002} proposed PROPEL for property Elucidation, and developed pattern templates for two representations in finite-state automata and a disciplined natural language. 
PSP is further extended with time by Gruhn et al.~\cite{Gruhn2006} for specifing real-time requirements, and probability and time in ProProST by Grunske~\cite{Grunske2008}.
Autili et al.~\cite{Autili2007} presents a scenario-based graphical language Property Sequence Chart (PSC) for the specification of temporal properties (LTL) in an extended UML sequence diagrams. The expressiveness of PSC is validated using Dwyer et al.'s PSP. Then Zhang et al. extended PSC with time in TPSC~\cite{Zhang2010} and with probability in PTPSC~\cite{Zhang2011}. They use the ProProST pattern system to meansure the expressiveness of PTPSC. 
Czepa et al.~\cite{Czepa2020} studied the understandability of three temporal property languages: LTL, PSP, and Event Processing Language~\cite{EsperTech2017}, and established that, comparatively, LTL is difficult to understand and PSP is easy to understand.
Different from these PSP-based approaches which are interpreted in either LTL, CTL or PCTL, our RoboCertProb is more expressive because it is based on PCTL*. It is slightly difficult to understand, compared to PSP, because RoboCertProb is a CNL and uses similar formula structures as PCTL*. To support PSP in RoboCertProb in the future would benefit roboticists.

In~\cite{Balasubramanian2011}, Balasubramanian et al. considered property specifications in model-based development where implementation code (in Java) for verification is automatically generated from abstract models captured in Simulink diagrams. It uses two approaches to specify properties for abstract models: specification patterns and contracts. These properties are then translated to properties for verification in Java Pathﬁnder~\cite{Visser2003}, a model checking tool for Java programs. Similarly, RoboCertProb is also developed using model-based techniques and used to specify property for abstract RoboChart models. Both RoboChart models and RoboCertProb will be automatically generated into models and properties for PRISM. RoboCertProb, however, supports both qualitative and quantitative property specification, while the specified properties in ~\cite{Balasubramanian2011} are only qualitative. 

In \cite{Zervoudakis2013}, Zervoudakis et al. proposed cascading veriﬁcation in which 
\begin{enumerate*}[label={(\arabic*)}]
    \item domain knowledge is modelled in the Web Ontology Language (OWL), the Semantic Web Rule Language (SWRL), and Prolog, 
    \item system behaviour is modelled in PRISM templates, and
    \item system property specifications are encoded in a high-level DSL, based on YAML\footnote{\url{https://yaml.org/}.} (a human-friendly data serialization language). 
\end{enumerate*}
A compiler then synthesises domain knowledge, models, and properties into DTMC models and PCTL properties for verification in PRISM. In our approach, we use the graphical notation RoboChart to capture system behaviour instead of PRISM in cascading verification, and so our approach is more accessible to engineers like roboticists. Our semantics supports both DTMC and MDP models. RoboCertProb also has a seamless integration with RoboChart models.  

In \cite{Sin2022}, a graphical notation TimeLinedepic is proposed to describe non-probabilistic temporal properties. It also presents a transformation from TimeLinedepic to the input property specification (LTL) for the SPIN model checker~\cite{Holzmann1997}. RoboCertProb is a textual notation. Its semantics is based on PCTL*, and so it is able to specify quantitative properties.

In \cite{Cardenas2023}, a new property specification technique is proposed to capture temporal property patterns PSP~\cite{Dwyer1999} in Temporal Object Constraint Language (TOCL)~\cite{Ziemann2004}. A validation tool is developed to support users with automatically verify UML class diagrams against the properties in TOCL. Unlike RoboCertProb, this work only supports the specification of non-probabilistic properties.

The work in~\cite{Barza2016} uses CNLs to write system requirements and capture temporal properties. They are automatically translated to models in the NuSMV model checker~\cite{Cimatti2002} and properties in CTL. Similarly, RoboCertProb is also a CNL. But RoboCertProb specifies properties for systems captured in RoboChart, a DSL, instead of a CNL used in~\cite{Barza2016} to capture system behaviours. Their property CNL is also not able to specify quantitative properties.

In \cite{Ghosh2016}, authors proposes ARSENAL to process temporal requirements in natural languages for safety critical systems into LTL. Comparatively, RoboCertProb is a CNL and specifies properties expressed in PCTL*.

The work~\cite{Santos2018} defines a CNL for specifying restrictions on the environment of a system. Derived from the restrictions, LTL formulae rule out infeasible scenarios in the CSP specification models according to the environmental restrictions. The techniques, in general, are used to generate test cases for CSP models from a CNL. RoboCertProb captures qualitative and quantitative properties in terms of PCTL*, and its expressive power is richer than~\cite{Santos2018}. Our language specifies properties for the DSL RoboChart, while the CNL~\cite{Santos2018} only specifies environmental restrictions.

In \cite{Vogel2023}, authors developed a property specification pattern catalogs (based on Timed CTL~\cite{Henzinger1994}) for UPPAAL to support qualitative and real-time requirements. RoboCertProb can specify both qualitative and quantitative properties, but not real-time properties.

RoboCertSeq~\cite{Windsor2022} is a property specification for RoboChart models with standard state machines and time features but without probabilistic features.
Properties specified in RoboCertSeq are qualitative, while RoboCertProb allows specifying both qualitative and quantitative properties. RoboCertSeq can specify time properties for the RoboChart models with the discrete-time feature, including time budgets, deadlines and timeouts, and RoboCertProb, however, cannot. Instead, RoboCertProb can specify time properties if time is modelled in RoboChart models as usual variables, such as \lstinbnf{ticks} in modelling the UV-light treatment robot, as discussed previously.

The PRISM property language\footnote{\url{www.prismmodelchecker.org/manual/PropertySpecification/}} supports property specification for discrete-time Markov models in PCTL* and continuous-time Markov models in CSL~\cite{ASSB96}. RoboCertProb is closely related to it and indeed inspired by it. RoboCertProb can be seen as a subset of PRISM's property language for discrete-time Markov models. While the PRISM property language targets users with knowledge of the PRISM language, the model checker, and the underlying temporal logics, RoboCertProb is designed for roboticists who use RoboChart for modelling and RoboCertProb to specify properties. We aim to provide a more ``natural'' way to specify properties. In addition to property specification, RoboCertProb provides facilities to instantiate RoboChart loose models and specify the environment.

To the best of our knowledge, RoboCertProb is the first comprehensive property specification CNL to 
\begin{enumerate*}[label={(\arabic*)}]
\item have PCTL* semantics, 
\item seamlessly integrate with DSLs, 
\item provide capabilities to instantiate loosely specified parametric models and constrain environment to support verifiction of reachability for particular inputs,
\item automate validation, cross-reference resolving, and code generation using model-based techniques.
\end{enumerate*}

\section{Conclusion and future work}
\label{sec:conclusion}

This work presents RoboCertProb, a qualitative and quantitative property specification for probabilistic RoboChart models. RoboCertProb's semantics is based on PCTL*. To interpret RoboCertProb over RoboChart models, we give RoboChart a Markov semantics in DTMCs and MDPs, derived from the existing RoboChart's PRISM semantics~\cite{Ye2022}. We present its syntax, associated well-formedness conditions, and design choices for the use of RoboCertProb to instantiate RoboChart models, specify environment and properties in terms of cross-referred RoboChart elements. We discuss its implementation in RoboTool through the Xtext framework for modelling, validation, scoping, and code generation. With the tool support, we can automatically generate PRISM properties from RoboCertProb. These properties then are analysed over the automatically generated PRISM models from the associate RoboChart models, as discussed in~\cite{Ye2022}. We use RoboTool to verify the properties of an industrial painting robot and analyse hazards for an agricultural UV-light treatment robot.

Our work can be extended to support the specification of broad properties such as steady-state behaviour, multi-objective properties, and filters that are supported in PRISM. 

We will enrich RoboCertProb's type and expression system to support those of RoboChart (based on the Z notation), and so we can specify properties using mathematical definitions like sets, relations, functions, sequences, and their corresponding operators. The introduction of these abstract mathematical types and expressions would facilitate the specification of properties for RoboChart models using RoboCertProb. To support such property specifications, we need to use data refinement~\cite{Woodcock1996} to refine these expressions into concrete expressions that probabilistic model checkers can support, or use theorem proving for verification.

The environment modelling in RoboCertProb through \lstinbnf{pModules} requires users to understand the PRISM language well, which is against the intention of RoboCertProb for roboticists. This can be mitigated by using RoboChart state machines to specify the environment, such as Fig.~\ref{fig:P3}, or extending sequence diagrams in RoboCertSeq with probability.


\vspace{1ex}

\noindent \textbf{Acknowledgements.}  This work is funded by the EPSRC grants EP/M025756/1 and EP/R025479/1. The icons used in RoboChart have been made by Sarfraz Shoukat, Freepik, Google, Icomoon and Madebyoliver from \url{www.flaticon.com} and are licensed under CC 3.0 BY.

We thank David A. Anisi and Mustafa Adam from the Norwegian University of Life Sciences for sharing the UV-light treatment robot case study, Yvonne Murray and Martin Sirev\r{a}g from the University of Agder for sharing the HVC case study with us.


\bibliographystyle{splncs}
\bibliography{main}


\ifdefined \CHANGES \indexprologue{%
  This index lists for each comment the pages where the text has been modified to address the comment. Since the same page may contain multiple changes, the page number contains the index of the change in superscript to identify different changes. Finally, the page number contains a hyperlink that takes the reader to the corresponding change.%
}%
\printindex[changes] \fi

\end{document}